\RequirePackage{fixltx2e}
\documentclass[preprint,10pt]{sigplanconf}
\usepackage{float}
\floatstyle{boxed}
\restylefloat{figure}

\usepackage{amssymb,amsmath,amsthm}
\allowdisplaybreaks
\usepackage{supertabular}
\usepackage{array}
\usepackage{etoolbox}
\usepackage{needspace}
\usepackage{url}
\usepackage{proof}
\usepackage[all]{xy}
\usepackage{tikz}
\usetikzlibrary{arrows,decorations.pathreplacing} 
\tikzset{
  to*/.style={
    shorten >=.25em,#1-to,
    to path={-- node[inner sep=0pt,at end,sloped] {${}^*$} (\tikztotarget) \tikztonodes}
  },
  to*/.default=
}

\makeatletter
\pgfarrowsdeclare{to*}{to*}
{
  \pgfutil@tempdima=-0.84pt%
  \advance\pgfutil@tempdima by-1.3\pgflinewidth%
  \pgfutil@tempdimb=0.21pt%
  \advance\pgfutil@tempdimb by.625\pgflinewidth%
  \advance\pgfutil@tempdimb by2.5pt%
  \pgfarrowsleftextend{+\pgfutil@tempdima}
  \pgfarrowsrightextend{+\pgfutil@tempdimb}
}
{
  \pgfutil@tempdima=0.28pt%
  \advance\pgfutil@tempdima by.3\pgflinewidth%
  \pgfsetlinewidth{0.8\pgflinewidth}
  \pgfsetdash{}{+0pt}
  \pgfsetroundcap
  \pgfsetroundjoin
  \pgfpathmoveto{\pgfqpoint{-3\pgfutil@tempdima}{4\pgfutil@tempdima}}
  \pgfpathcurveto
  {\pgfqpoint{-2.75\pgfutil@tempdima}{2.5\pgfutil@tempdima}}
  {\pgfqpoint{0pt}{0.25\pgfutil@tempdima}}
  {\pgfqpoint{0.75\pgfutil@tempdima}{0pt}}
  \pgfpathcurveto
  {\pgfqpoint{0pt}{-0.25\pgfutil@tempdima}}
  {\pgfqpoint{-2.75\pgfutil@tempdima}{-2.5\pgfutil@tempdima}}
  {\pgfqpoint{-3\pgfutil@tempdima}{-4\pgfutil@tempdima}}
  \pgfusepathqstroke
  \begingroup
  \pgftransformxshift{2.5pt}
  \pgftransformyshift{2pt}
  \pgftransformscale{.7}
  \pgfuseplotmark{asterisk}
\endgroup
}

\pgfarrowsdeclare{*to}{*to}
{
  \pgfutil@tempdima=-0.84pt%
  \advance\pgfutil@tempdima by-1.3\pgflinewidth%
  \pgfutil@tempdimb=0.21pt%
  \advance\pgfutil@tempdimb by.625\pgflinewidth%
  \advance\pgfutil@tempdimb by2.5pt%
  \pgfarrowsleftextend{+\pgfutil@tempdima}
  \pgfarrowsrightextend{+\pgfutil@tempdimb}
}
{
  \pgfutil@tempdima=0.28pt%
  \advance\pgfutil@tempdima by.3\pgflinewidth%
  \pgfsetlinewidth{0.8\pgflinewidth}
  \pgfsetdash{}{+0pt}
  \pgfsetroundcap
  \pgfsetroundjoin
  \pgfpathmoveto{\pgfqpoint{-3\pgfutil@tempdima}{4\pgfutil@tempdima}}
  \pgfpathcurveto
  {\pgfqpoint{-2.75\pgfutil@tempdima}{2.5\pgfutil@tempdima}}
  {\pgfqpoint{0pt}{0.25\pgfutil@tempdima}}
  {\pgfqpoint{0.75\pgfutil@tempdima}{0pt}}
  \pgfpathcurveto
  {\pgfqpoint{0pt}{-0.25\pgfutil@tempdima}}
  {\pgfqpoint{-2.75\pgfutil@tempdima}{-2.5\pgfutil@tempdima}}
  {\pgfqpoint{-3\pgfutil@tempdima}{-4\pgfutil@tempdima}}
  \pgfusepathqstroke
  \begingroup
  \pgftransformxshift{2.5pt}
  \pgftransformyshift{-2pt}
  \pgftransformscale{.7}
  \pgfuseplotmark{asterisk}
\endgroup
}

\makeatother

\newtheorem{theorem}{Theorem}[section]
\AtBeginEnvironment{theorem}{\Needspace{2\baselineskip}}
\newtheorem{lemma}[theorem]{Lemma}
\AtBeginEnvironment{lemma}{\Needspace{2\baselineskip}}
\theoremstyle{definition}
\newtheorem{definition}[theorem]{Definition}
\AtBeginEnvironment{definition}{\Needspace{2\baselineskip}}
\newtheorem{example}[theorem]{Example}
\AtBeginEnvironment{example}{\Needspace{2\baselineskip}}

\newcommand\citaLemaOrig[1]{\cite[Lemma #1]{DiazcaroDowek15}}
\newcommand\citaTeoOrig[1]{\cite[Theorem #1]{DiazcaroDowek15}}
\newcommand\eq{\ensuremath{\rightleftarrows}}
\newcommand\re{\ensuremath{\hookrightarrow}}
\newcommand\toreq{\ensuremath{\rightsquigarrow}}
\newcommand\ve[1]{\ensuremath{\mathrm{\bf #1}}}
\newcommand\Red[1]{\ensuremath{\mathrm{Red}({#1})}}
\newcommand\s[1]{\ensuremath{\mathsf{#1}}}
\newcommand\can[1]{\ensuremath{\s{can}(#1)}}
\newcommand\uncan[1]{\ensuremath{\s{nac}(#1)}}
\newcommand\canv[1]{\ensuremath{\s{can}\!\left(\ve{#1}\right)}}

\newcommand\ms[4][1]{\ensuremath{[#2]_{#3=#1}^{#4}}} % \ms[2]{C_i}{i}{n} == [C_i]_{i=1}^n
\newcommand\ws[4][1]{\ensuremath{\bigwedge_{#3=#1}^{#4}{#2}}} 
\newcommand\wsl[4][1]{\ensuremath{\bigwedge\limits_{#3=#1}^{#4}{#2}}} 
\newcommand\sms[1]{\ensuremath{[#1]}}
\newcommand\gms[1]{\ensuremath{[\vec{#1}]}}
\newcommand\conlista{\leavevmode\vspace{-0.4\baselineskip}}

\newcommand\cond[2][A]{\mbox{\scriptsize$\deduce{\phantom{#1}}{(#2)}$}\ }

\newcommand\suc[1]{\ensuremath{\s{succ}~{#1}}}
\newcommand\pred[1]{\ensuremath{\s{pred}~{#1}}}
\newcommand\num[1]{\ensuremath{\mathfrak{#1}}}
\newcommand\bnum[1]{\ensuremath{\bar{\num{#1}}}}
\newcommand\ifZ{\ensuremath{\s{ifZ}}}
\newcommand\ifEq{\ensuremath{\s{ifEq}}}
\newcommand\fst{\ensuremath{\s{fst}_{Nat}}}
\newcommand\fstO[1]{\ensuremath{\s{fst}_{#1}}}
\newcommand\snd{\ensuremath{\s{snd}_{Nat}}}
\newcommand\sndO[1]{\ensuremath{\s{snd}_{#1}}}
\newcommand\divModRec[2]{\ensuremath{\s{divModRec}^{\num{#1}\num{#2}}}}
\newcommand\divMod[2]{\ensuremath{\s{divMod}^{\num{#1}\num{#2}}}}
\newcommand\evenOdd{\ensuremath{\s{evenOdd}}}
\newcommand\swap{\ensuremath{\s{swap}}}
\newcommand\succFst{\ensuremath{\s{succFst}}}

\newcommand\canon[2]{\ensuremath{[{#1}]^{\bnum{#2}}}}
\newcommand\estr[1]{\ensuremath{\star^{\num{#1}}}}
\newcommand\cnat[1]{\canon{Nat}{#1}}

\newcommand\terrulelabel[1]{\mbox{\scriptsize\sc{#1}}}

\newcommand\repl[1]{\{#1\}}
\newcommand\tf{\mbox{\bf T\!F}}
\newcommand\true{\mbox{\bf T}}
\newcommand\false{\mbox{\bf F}}
\newcommand\m[2]{\multicolumn{#1}{>{\(}l<{\)}}{#2}}
\newcommand\expl[2]{\m{50}{#1\ \mbox{\small\textit{(#2)}}}}
\newcommand\all[1]{\m{50}{#1}}
\newcommand{\xrecap}[2]{\noindent\textbf{#1 \ref{#2}.}}

\begin{document}

\setlength{\pdfpageheight}{\paperheight}
\setlength{\pdfpagewidth}{\paperwidth}

\conferenceinfo{IFL'15}{September 14--16, 2015, Koblenz, Germany} 
\copyrightyear{2015} 
\copyrightdata{978-1-nnnn-nnnn-n/yy/mm} 
\doi{nnnnnnn.nnnnnnn}

% Uncomment one of the following two, if you are not going for the 
% traditional copyright transfer agreement.

%\exclusivelicense                % ACM gets exclusive license to publish, 
				  % you retain copyright

%\permissiontopublish             % ACM gets nonexclusive license to publish
				  % (paid open-access papers, 
				  % short abstracts)

%\titlebanner{banner above paper title}        % These are ignored unless
%\preprintfooter{short description of paper}   % 'preprint' option specified.

\title{Isomorphisms considered as equalities} 
\subtitle{Projecting functions and enhancing partial application through an implementation of $\lambda^+$}
%An implementation of simply typed lambda-calculus modulo type isomorphisms}
%\subtitle{Subtitle Text, if any}

\authorinfo{Alejandro D\'iaz-Caro \and Pablo E.~Mart\'inez L\'opez}
{Universidad Nacional de Quilmes}
{\tt alejandro.diaz-caro@unq.edu.ar \and fidel@unq.edu.ar}

\maketitle

\begin{abstract}
  We propose an implementation of $\lambda^+$, a recently introduced simply typed lambda-calculus with pairs where isomorphic types are made equal. The rewrite system of $\lambda^+$ is a rewrite system modulo an equivalence relation, which makes its implementation non-trivial.
  We also extend $\lambda^+$ with natural numbers and general recursion and use Beki\'c's theorem to split mutual recursions. This splitting, together with the features of $\lambda^+$, allows for a novel way of program transformation by reduction, by projecting a function before it is applied in order to simplify it. Also, currying together with the associativity and commutativity of pairs gives an enhanced form of partial application.
\end{abstract}

\category{F.4.1}{Mathematical Logic}{Lambda calculus and related systems}

% general terms are not compulsory anymore, 
% you may leave them out
%\terms
%term1, term2

\keywords
$\lambda$-calculus, type isomorphisms, implementation

\section{Introduction}
From the study of non-determinism in quantum computing~\cite{ArrighiDiazcaroLMCS12,ArrighiDiazcaroValiron13} a non-deterministic extension of the simply typed lambda calculus with conjunction called $\lambda^+$ has been introduced~\cite{DiazcaroDowek15}. The non-determinism in this calculus arises by taking the type isomorphisms as equalities to generate a type system modulo such an equivalence relation. Since $R\wedge S$ is isomorphic to $S\wedge R$, in a type system modulo isomorphisms the pair construction is impervious of the order of its constituents. This way, $\langle\ve r,\ve s\rangle$ is equal to $\langle\ve s,\ve r\rangle$ and hence a projection over a pair cannot depend on the position. Instead of having $\pi_1$ and $\pi_2$ as projectors, in $\lambda^+$ the projection depends on the type, so if $\ve r$ has type $R$, the projection of $\langle\ve r,\ve s\rangle$ with respect to such a type would reduce to $\ve r$. Because of the commutativity of pairs, in the case $\ve s$ has also type $R$, the projection behaves non-deterministically projecting either $\ve r$ or $\ve s$. Pairs in $\lambda^+$ are noted with ``$+$'' to emphasise their non-deterministic nature.

With simply types only four isomorphisms are enough to characterise any other isomorphism in the system~\cite{BruceDiCosmoLongoMSCS92}:
\begin{align*}
  R\wedge S&\equiv S\wedge R & \mbox{\small(comm)}\\
  (R\wedge S)\wedge T &\equiv R\wedge (S\wedge T) &\mbox{\small(assoc)}\\
  R\Rightarrow (S\wedge T)&\equiv (R\Rightarrow S)\wedge (R\Rightarrow T) &\mbox{\small(dist)}\\
  R\Rightarrow S\Rightarrow T &\equiv (R\wedge S)\Rightarrow T &\mbox{\small(curry)}
\end{align*}

Although the main aim of $\lambda^+$ has been to study the non-determinism in quantum computing, the novelty of a type system modulo type isomorphisms lies in its many good properties. In particular, isomorphism (dist) implies that a function returning two arguments is the same as a pair of functions, and so it can be projected, even before the function is calculated (see examples in Section~\ref{sec:example}). In addition, the isomorphism (curry) together with (comm) and (assoc) induces a system where functions returning functions are just functions with more parameters, and where the order these parameters are given is irrelevant. This allows for an enhanced form of partial application.

In Functional Programming, partial application allows defining the successor function in the following way:
\begin{align*}
  \s{addition} &= \lambda x^{Nat}.\lambda y^{Nat}.(x+y)\\
  \s{succ} &= \s{addition}~1
\end{align*}
That is, the successor function is expressed as a partial application of the addition function. Our system modulo isomorphisms allows also to define it in the following way:
\begin{align*}
  \s{addition}' &= \lambda x^{Nat\times Nat}.(\s{fst}(x)+\s{snd}(x))\\
  \s{succ}' &= \s{addition}'~1
\end{align*}
(with some encoding of $\s{fst}$ and $\s{snd}$ ensuring that they do not behave non-deterministically on pairs, cf.~Section~\ref{sec:nat}).

Another, more interesting example, is that not only the partial application can occur by passing the first argument, but any argument can be passed before passing the remaining ones. For example, assume we are given a function \s{elem} which receives an element and a list of elements and checks whether such an element is in the list. Then, we could define a function to check if a given color is a primary color as $\s{isPrimary} = \s{elem}\ {\s{list\_of\_primary\_colors}}$. Note that \s{elem} has been applied to a list rather than an element.
%owever, since \s{elem} is expecting to receive first the element and then the list, a flip function would be needed. 
When isomorphisms are considered as equalities, we have
\begin{align*}
  R\Rightarrow S\Rightarrow T &\equiv (R\wedge S)\Rightarrow T\\
  & \equiv (S\wedge R)\Rightarrow T\\
  &\equiv S\Rightarrow R\Rightarrow T
\end{align*}
Hence,
%and so no transformation need to be done to \s{elem}, 
passing any argument first will not change its typability. %the implementation.

At a theoretical level, $\lambda^+$ has several interesting advantages; however, implementing a type system with a modulo structure is not straightforward, and is thus the goal of this paper.
In order to have a meaningful  language, we introduce natural numbers and general recursion. As stated before, by the (dist) isomorphism, a function returning pairs can be considered as a pair of functions, which can be projected. Hence, we do the same with the general recursion using Beki\'c's theorem~\cite{BekicLNCS}. This allows us to define, with mutual recursion, functions returning two values and then to project one of them. Only that instead of doing all the calculation and then projecting the needed result, our system allows projecting the needed pieces of code from the function, so the function would be optimized to do only the wanted calculation.

A prototype of this implementation has been written in Haskell. The sources can be found at 
\begin{center}
  {\small \url{http://www.diaz-caro.info/IsoAsEq-v1.0.tar.gz}}
\end{center}

\subsection*{Plan of the paper} 
Section~\ref{sec:original} presents $\lambda^+$ in brief (please, refer to~\cite{DiazcaroDowek15} for a more detailed description). Section~\ref{sec:types} details the implementation of isomorphic types. Section~\ref{sec:rewrite} does the same with the rewrite relation modulo. Section~\ref{sec:nat} presents an encoding for tuples and extends the language with natural numbers and general recursion. Section~\ref{sec:example} shows two interesting examples of projecting recursive functions. Finally, Section~\ref{sec:conclusion} concludes and proposes some future work.
The proofs of lemmas are left out of the paper due to length restrictions. The reviewer can find all the proofs in the appendix on this draft version for his convenience.

\section{The Original Setting}\label{sec:original}
In this section we present a brief description of $\lambda^+$. 
%For more details the reader is invited to refer to~\cite{DiazcaroDowek15}.
The grammar of types is given by
\[
  R,S ::= \tau~|~R\Rightarrow S~|~R\wedge S
\]
where $\tau$ is an atomic type.
We use upper case letters such as $R$, $S$, and $T$ for types.

The type isomorphisms mentioned in the introduction, (comm), (assoc), (dist) and (curry), are taken as a congruent equivalent relation over types, denoted by $\equiv$.

The grammar of terms is the following
\[
  \ve r, \ve s::=\quad x^R~|~\lambda x^R.\ve r~|~\ve r\ve s~|~\ve r+\ve s~|~\pi_R(\ve r)
\]

We use bold lower case letters such as $\ve r$, $\ve s$, $\ve t$ for generic terms, and normal lower case letters such as $x$, $y$, $z$ for variables.

A context is a finite set of typed variables, such as $\Gamma=x^R,y^S,z^T$. We use upper case Greek letters such as $\Gamma$ and $\Delta$ for contexts. The same variable cannot appear twice with different types in a context.

The set of term contexts is defined inductively by the grammar
\begin{align*}
  C[\cdot] & ::= [\cdot]~|~\lambda x^R.C[\cdot]~|~C[\cdot]\ve r~|~\ve rC[\cdot]\\
  &\ \  |~C[\cdot]+\ve r~|~\ve r+C[\cdot]~|~\pi_R(C[\cdot])
\end{align*}

The type system is given in Figure~\ref{fig:origTypes}. Typing judgements are of the form $\Gamma\vdash\ve r:R$. A term $\ve r$ is typable if there exists a type $R$ and a set of typed variables $\Gamma$ such that $\Gamma\vdash\ve r:R$. 
\begin{figure*}[!t]
  \[
    \infer[^{(\textsl{ax})}]{\Gamma,x^R\vdash x:R}{\phantom{x^R}}
    \quad
    {\cond{R\equiv R'}}\infer[^{(\equiv)}]{\Gamma\vdash\ve r:R}{\Gamma\vdash\ve r:R'}
    \quad
    \infer[^{(\Rightarrow_i)}]
    {\Gamma\vdash\lambda x^S.\ve r:S\Rightarrow R}
    {\Gamma,x^S\vdash\ve r:R}
    \quad
    \infer[^{(\Rightarrow_e)}]
    {\Gamma\vdash\ve r\ve s:R}
    {\Gamma\vdash\ve r:S\Rightarrow R & \Gamma\vdash\ve s:S}
  \]
  \[
    \infer[^{(\wedge_i)}]
    {\Gamma\vdash\ve r+\ve s:R\wedge S}
    {\Gamma\vdash\ve r:R & \Gamma\vdash\ve s:S}
    \qquad
    \infer[^{(\wedge_{e_n})}]{\Gamma\vdash\pi_R(\ve r):R}{\Gamma\vdash\ve r:R\wedge R'}
    \qquad
    \infer[^{(\wedge_{e_1})}]{\Gamma\vdash\pi_R(\ve r):R}{\Gamma\vdash\ve r:R}
  \]
  \caption{The $\lambda^+$ type system}
  \label{fig:origTypes}
\end{figure*}

Each variable occurrence is labelled by its type, such as $\lambda x^R.x^R$ or $\lambda x^R.y^S$. We sometimes omit the labels when it is clear from the context, for example we may write $\lambda x^R.x$ for $\lambda x^R.x^R$ or $x^R\vdash x:R$ for $x^R\vdash x^R:R$. As usual, we consider implicit $\alpha$-equivalence on syntactical terms. 
Because of the condition on contexts where the same variable cannot appear twice with different types in a context, the type system forbids terms such as $\lambda x^R.x^S$ when $R$ and $S$ are not equivalent types.

The set $FV(\ve r)$ of free variables of $\ve r$ is defined as expected.
For example, the set $FV(\lambda x^{R\Rightarrow S\Rightarrow T}.xy^Rz^S)$ is $\{y^R,z^S\}$.
We say that a term $\ve t$ is closed whenever $FV(\ve r)=\emptyset$.

Given two terms $\ve r$ and $\ve s$ we denote by $\ve r\repl{\ve s/x}$ the term obtained by simultaneously substituting the term $\ve s$ for all the free occurrences of $x$ in $\ve r$, subject to the usual proviso about renaming bound variables in $\ve r$ to avoid capture of the free variables of $\ve s$. 

Lemma~\ref{lem:unicity} states that the type system assigns a unique type to each term, modulo isomorphisms.
\begin{lemma}[\citaLemaOrig{2.1}]\label{lem:unicity}~

  If $\Gamma\vdash\ve r:R$ and $\Gamma\vdash\ve r:R'$, then $R\equiv R'$.
  \qed
\end{lemma}

The operational semantics of the calculus is given in Figure~\ref{fig:opSemOrig}, where there are two distinct relations between terms: a symmetric relation $\eq$ and a reduction relation $\re$, which includes a labelling $\neg\delta$ or $\delta$. Such a labelling is omitted when it is not necessary to distinguish the rule. Moreover, relation $\re$ is $\stackrel{\neg\delta}{\re}\cup\stackrel{\delta}{\re}$. 
It is used only to distinguish rule $(\delta)$ from the other rules, as it is standard with surjective pairing expansion \cite{DicosmoKesnerLNCS93}.
Rule \terrulelabel{($\delta$)} has been added to deal with curryfication, (cf.~Example~\ref{ex:delta}).
The conditions on this rule are standard for surjective pairing in the expansive direction in order to avoid cycling.
Type substitution on a term $\ve r$, written $\ve r\repl{S/R}$, is defined as the syntactic substitution of all occurrences of the type $R$ in $\ve r$ by $S$.
We write $\re^*$ and $\eq^*$ for the transitive and reflexive closure of $\re$ and $\eq$ respectively. Note that $\eq^*$ is an equivalence relation.

Let  $\toreq$ be the relation $\re$ modulo $\eq^*$ (i.e. $\ve r\toreq\ve s$ iff $\ve r\eq^*\ve r'\re\ve s'\eq^*\ve s$), and $\toreq^*$ its reflexive and transitive closure.
We say that two terms $\ve r$ and $\ve s$ are observationally equivalent if they can be typed by the same type, and for each $\ve t$ such that $\ve r\ve t$ and $\ve s\ve t$ are well typed, there exists $\ve t'$ such that $\ve r\ve t\toreq^*\ve t'$ and $\ve s\ve t\toreq^*\ve t'$.
\begin{figure*}[!t]
  \[
    \begin{array}{rcl@{\qquad}r}
      \multicolumn{4}{l}{\mbox{\bf Symmetric relation:}}\\
      \ve r+\ve s &\eq& \ve s+\ve r & \terrulelabel{(comm)}\\
      (\ve r+\ve s)+\ve t &\eq& \ve r+(\ve s+\ve t) & \terrulelabel{(asso)}\\
      \lambda x^R.(\ve r+\ve s) &\eq& \lambda x^R.\ve r+\lambda x^R.\ve s & \terrulelabel{(dist$_{ii}$)}\\
      (\ve r+\ve s)\ve t &\eq& \ve r\ve t+\ve s\ve t & \terrulelabel{(dist$_{ie}$)}\\
      \pi_{R\Rightarrow S}(\lambda x^R.\ve r) & \eq &\lambda x^R.\pi_S(\ve r) & \terrulelabel{(dist$_{ei}$)}\\
      \mbox{If }
      \left\lbrace\begin{array}{l}
	\Gamma\vdash\ve r:R\Rightarrow (S\wedge T)\\
	\Gamma\vdash\ve s:R\\
      \end{array}\right\rbrace,\ 
      \pi_{R\Rightarrow S}(\ve r)\ve s &\eq &\pi_S(\ve r\ve s) & \terrulelabel{(dist$_{ee}$)}\\
      \ve r\ve s\ve t &\eq& \ve r(\ve s+\ve t) & \terrulelabel{(curry)}\\ 
      \text{If }R\equiv S,\ \ve r &\eq& \ve r\repl{S/R} & \terrulelabel{(subst)}\\
      \text{If }
      \left\lbrace\begin{array}{l}
	\Gamma\vdash\ve r:R\wedge R'\mbox{ or }\Gamma\vdash\ve r:R\\\
	\Gamma\vdash\ve s:S\wedge S'\mbox{ or }\Gamma\vdash\ve s:S\\
      \end{array}\right\rbrace,\ 
      \pi_{R\wedge S}(\ve r+\ve s)&\eq&\pi_R(\ve r)+\pi_S(\ve s) &\terrulelabel{(split)}\\
      \\
      \multicolumn{4}{l}{\mbox{\bf Reductions:}}\\
    \mbox{If }\Gamma\vdash\ve s:R,\  (\lambda x^R.\ve r)\ve s &\stackrel{\neg\delta}{\re}& \ve r\repl{\ve s/x} & \terrulelabel{($\beta$)}\\
      \mbox{If }\Gamma\vdash\ve r:R,\ \pi_R(\ve r+\ve s) &\stackrel{\neg\delta}{\re}& \ve r & \terrulelabel{($\pi_n$)}\\
      \mbox{If }\Gamma\vdash\ve r:R,\ \pi_R(\ve r) &\stackrel{\neg\delta}{\re}& \ve r & \terrulelabel{($\pi_1$)}\\
      \mbox{If }
      \left\lbrace\begin{array}{l}
	\ve \Gamma\vdash\ve r:R\wedge S,\\
	\ve r\not\eq^*\ve s+\ve t \mbox{ with }
	\Gamma\vdash\ve s:R\mbox{ and }\Gamma\vdash\ve t:S
      \end{array}\right\rbrace,\ 
      \ve r &\stackrel{\delta}{\re}&\pi_R(\ve r)+\pi_S(\ve r) & \terrulelabel{($\delta$)}
    \end{array}
  \]
  \[
    \infer[^{(C_\eq)}]{C[\ve r]\eq C[\ve s]}{\ve r\eq\ve s}
    \qquad
    \infer[^{(C_{\re}^{(\neg\delta)})}]{C[\ve r]\stackrel{\neg\delta}{\re} C[\ve s]}{\ve r\stackrel{\neg\delta}{\re}\ve s}
    \qquad
    \infer[^{(C_{\re}^{\delta})}]{C[\ve r]\stackrel{\delta}{\re}C{[\ve s]}}
    {
      \ve r\stackrel{\delta}{\re}\ve s
      &
      C[\cdot]\not\eq^*C'[\pi_R(\cdot)]
    }
  \]
  \caption{The $\lambda^+$ operational semantics} 
  \label{fig:opSemOrig}
\end{figure*}

Each isomorphism taken as an equivalence between types induces an equivalence between terms, given by relation $\eq$. 
Four possible rules exist however for the isomorphism~(dist), depending of which distribution is taken into account: elimination or introduction of conjunction, and elimination or introduction of implication. 

Only two rules in the symmetric relation $\eq$ are not a direct consequence of an isomorphism: rules \terrulelabel{(subst)} and \terrulelabel{(split)}. The former allows updating the type annotations of the Church-style terms. The latter is needed to be used in combination with rule \terrulelabel{(dist$_{ei}$)} when the argument in the projection is not a $\lambda$-abstraction, but a $\lambda$-abstraction plus something else (cf.~Example~\ref{ex:split}).

Lemma~\ref{lem:finiteClasses} ensures that the equivalence classes defined by relation $\eq^*$,
$\{\ve s~|~\ve s \eq^* \ve r\}$, are finite, and since the relation is 
computable, the side condition of \terrulelabel{($\delta$)} is decidable.
In addition, Lemma \ref{lem:finiteClasses} implies that every $\re$-reduction modulo $\eq^*$ tree is finitely branching.

\begin{lemma}[\citaLemaOrig{2.4}]\label{lem:finiteClasses}~

  For any term $\ve r$, the set $\{\ve s~|~\ve s \eq^*\ve r\}$ is finite (modulo
  $\alpha$-equivalence).
  \qed
\end{lemma}

Notice that because of the commutativity and associativity properties of $\wedge$, the symbol $+$ on terms is also taken as commutative and associative. Hence, the term $\ve r+(\ve s+\ve t)$ is the same as the term $(\ve r+\ve s)+\ve t$, so it can be expressed just as $\ve r+\ve s+\ve t$, that is, the parenthesis are meaningless, and pairs become multisets. In particular, we can project with respect to the type of a sum. This is why, for completeness, we also allow projecting a term with respect to its full type, that is, if $\Gamma\vdash\ve r:R$, then $\pi_R(\ve r)$ reduces to $\ve r$.

\subsection{Examples}
\begin{example}
  Let $\vdash\ve r:R$ and $\vdash\ve s:S$. Then

  \hspace{-5mm}\scalebox{0.83}{
    \parbox{10cm}{
      \[
	\infer[^{(\Rightarrow_e)}]{\vdash\pi_{S\Rightarrow R}((\lambda x^{R\wedge S}.x)\ve r)\ve s:R}
	{
	  \infer[^{(\wedge_{e_n})}]{\vdash\pi_{S\Rightarrow R}((\lambda x^{R\wedge S}.x)\ve r):S\Rightarrow R}
	  {
	    \infer[^{(\equiv)}]{\vdash(\lambda x^{R\wedge S}.x)\ve r:(S\Rightarrow R)\wedge(S\Rightarrow S)}
	    {
	      \infer[^{(\Rightarrow_e)}]{\vdash(\lambda x^{R\wedge S}.x)\ve r:S\Rightarrow(R\wedge S)}
	      {
		\infer[^{(\equiv)}]{\vdash\lambda x^{R\wedge S}.x:R\Rightarrow S\Rightarrow(R\wedge S)}
		{
		  \infer[^{(\Rightarrow_i)}]{\vdash\lambda x^{R\wedge S}.x:(R\wedge S)\Rightarrow(R\wedge S)}
		  {
		    \infer[^{(\textsl{ax})}]{x^{R\wedge S}\vdash x:R\wedge S}{}
		  }
		}
		&
		\vdash\ve r:R
	      }
	    }
	  }
	  &
	  \vdash\ve s:S
	}
      \]
    }
  }

  The reduction is as follows:
  \begin{align*}
    \pi_{S\Rightarrow R}((\lambda x^{R\wedge S}.x)\ve r)\ve s
    &\eq\pi_R((\lambda x^{R\wedge S}.x)\ve r\ve s)\\
    &\eq\pi_R((\lambda x^{R\wedge S}.x)(\ve r+\ve s))\\
    &\re\pi_R(\ve r+\ve s)\\
    &\re\ve r
  \end{align*}
\end{example}
\begin{example}
  Let $\vdash\ve r:R$, $\vdash\ve s:S$. Then
  \[
    (\lambda x^{R}.\lambda y^S.x)(\ve r+\ve s)
    \eq (\lambda x^{R}.\lambda y^S.x)\ve r\ve s
    \re^* \ve r
  \]
  However, if $R\equiv S$, it is also possible to reduce it in the following way

  \begin{align*}
    (\lambda x^{R}.\lambda y^S.x)(\ve r+\ve s)
    &\eq(\lambda x^{R}.\lambda y^R.x)(\ve r+\ve s)\\
    &\eq (\lambda x^{R}.\lambda y^R.x)(\ve s+\ve r)\\
    &\eq  (\lambda x^{R}.\lambda y^R.x)\ve s\ve r\\
    &\re^* \ve s
  \end{align*}
  Hence, the encoding of the projector also behaves non-deterministically.
\end{example}

\begin{example} Let $\tf=\lambda x^R.\lambda y^S.(x+y)$. Then
  \[
    \infer[^{(\wedge_e)}]{\vdash\pi_{R\Rightarrow S\Rightarrow R}(\tf):R\Rightarrow S\Rightarrow R}
    {
      \infer[^{(\equiv)}]{\vdash\tf:(R\Rightarrow S\Rightarrow R)\wedge(R\Rightarrow S\Rightarrow S)}
      {
	\infer[^{(\Rightarrow_i)}]{\vdash\tf:R\Rightarrow S\Rightarrow (R\wedge S)}
	{
	  \infer[^{(\Rightarrow_i)}]{x^R\vdash\lambda y^S.(x+y):S\Rightarrow(R\wedge S)}
	  {
	    \infer[^{(\wedge_i)}]{x^R,y^S\vdash x+y:R\wedge S}
	    {
	      \infer[^{(\textsl{ax})}]{x^R,y^S\vdash x:R}{}
	      &
	      \infer[^{(\textsl{ax})}]{x^R,y^S\vdash y:S}{}
	    }
	  }
	}
      }
    }
  \]
  Hence,
  if $\Gamma\vdash\ve r:R$ and $\Gamma\vdash\ve s:S$, we have
  \[
    \Gamma\vdash\pi_{R\Rightarrow S\Rightarrow R}(\tf)\ve r\ve s:R
  \]

  Notice that
  \begin{align*}
    \pi_{R\Rightarrow S\Rightarrow R}(\tf)\ve r\ve s
    &\eq\pi_{S\Rightarrow R}(\tf\ve r)\ve s\\
    &\eq    \pi_{R}(\tf\ve r\ve s)\\
    &\re^*    \pi_R(\ve r+\ve s)\\
    &\re   \ve r
  \end{align*}
\end{example}

\begin{example}\label{ex:bool}
  Let $\true=\lambda x^R.\lambda y^S.x$ and $\false=\lambda x^R.\lambda y^S.y$. Then 
  \(
  \vdash\true+\false+\tf:((R\Rightarrow S\Rightarrow R)\wedge(R\Rightarrow S\Rightarrow S))\wedge\left((R\Rightarrow S\Rightarrow R)\wedge (R\Rightarrow S\Rightarrow S)\right) 
  \).
  Therefore, the term
  \(
    \pi_{(R\Rightarrow S\Rightarrow R)\wedge(R\Rightarrow S\Rightarrow S)}(\true+\false+\tf)
  \) 
  is typable and reduces non-deterministically either to $\true+\false$ or to $\tf$. Moreover, notice that $\true+\false$ and $\tf$ are observationally equivalent, that is, $(\true+\false)\ve r\ve s$ and $\tf\ve r\ve s$ both reduce to the same term ($\ve r+\ve s$). 
%  The reduction diagram can be expressed as follows
%  \[
%    \xymatrix@C=2ex@R=3ex{
%      & \pi_{(R\Rightarrow S\Rightarrow R)\wedge(R\Rightarrow S\Rightarrow S)}(\true+\false+\tf) \ar@{^{(}->}[dr] \ar@{_{(}->}[dl]& \\
%      \true+\false \ar@{~~}[rr] & & \tf
%    }
%  \]
  Hence,
  in this very particular case, the non-deterministic choice does not play any role.
\end{example}

\begin{example}
  \label{ex:delta}
  Let $\vdash\ve r:T$. Then
  $\lambda x^{(R\wedge S)\Rightarrow R}.\lambda y^{(R\wedge S)\Rightarrow S}.\ve r$ has type $((R\wedge S)\Rightarrow R)\Rightarrow((R\wedge S)\Rightarrow S)\Rightarrow T$,
  and since $((R\wedge S)\Rightarrow R)\Rightarrow((R\wedge S)\Rightarrow S)\Rightarrow T\equiv((R\wedge S)\Rightarrow(R\wedge S))\Rightarrow T$, we also can derive
  \[
    \vdash\lambda x^{(R\wedge S)\Rightarrow R}.\lambda y^{(R\wedge S)\Rightarrow S}.\ve r:((R\wedge S)\Rightarrow(R\wedge S))\Rightarrow T
  \]
  Therefore,
  \[
    \vdash(\lambda x^{(R\wedge S)\Rightarrow R}.\lambda y^{(R\wedge S)\Rightarrow S}.\ve r)(\lambda z^{R\wedge S}.z):T
  \]
  The reduction occurs as follows: 
  \[
    \begin{array}{l}
      (\lambda x^{(R\wedge S)\Rightarrow R}.\lambda y^{(R\wedge S)\Rightarrow S}.\ve r)(\lambda z^{R\wedge S}.z)\\
      \re\ (\terrulelabel{$\delta$}) \\
      (\lambda x^{(R\wedge S)\Rightarrow R}.\lambda y^{(R\wedge S)\Rightarrow S}.\ve r)\\
      \left(\pi_{(R\wedge S)\Rightarrow R}(\lambda z^{R\wedge S}.z)
      +
      \pi_{(R\wedge S)\Rightarrow S}(\lambda z^{R\wedge S}.z)\right)\\
      \eq\ (\terrulelabel{curry})\\
      \left((\lambda x^{(R\wedge S)\Rightarrow R}.\lambda y^{(R\wedge S)\Rightarrow S}.\ve r)
      \pi_{(R\wedge S)\Rightarrow R}(\lambda z^{R\wedge S}.z)\right)\\
      \pi_{(R\wedge S)\Rightarrow S}(\lambda z^{R\wedge S}.z)\\
      \re^*\ (\terrulelabel{$\beta^{\times 2}$})\\
      \ve r
      \repl{\pi_{(R\wedge S)\Rightarrow R}(\lambda z^{R\wedge S}.z)/x}
      \repl{\pi_{(R\wedge S)\Rightarrow S}(\lambda z^{R\wedge S}.z)/y}
    \end{array}
  \]
\end{example}
\begin{example}
  \label{ex:split}
  Let $\vdash\ve r:T$. Then

  \hspace{-10mm}\scalebox{0.88}{
    \parbox{10cm}{
      \[
	\infer[^{(\wedge_e)}]{\vdash\pi_{((R\wedge S)\Rightarrow R)\wedge T}((\lambda x^{R\wedge S}.x)+\ve r):(     (R\wedge S)\Rightarrow R)\wedge T}
	{
	  \infer[^{(\equiv)}]{\vdash(\lambda x^{R\wedge S}.x)+\ve r:(((R\wedge S)\Rightarrow R)\wedge T)\wedge (R\wedge S)\Rightarrow S}
	  {
	    \infer[^{(\wedge_i)}]{\vdash(\lambda x^{R\wedge S}.x)+\ve r:(R\wedge S)\Rightarrow(R\wedge S)\wedge T}
	    {
	      \infer[^{(\Rightarrow_i)}]{\vdash\lambda x^{R\wedge S}.x:(R\wedge S)\Rightarrow(R\wedge S)}
	      {
		\infer[^{(\textsl{ax})}]{x^{R\wedge S}\vdash x:R\wedge S}{}
	      }
	      &
	      \vdash\ve r:T
	    }
	  }
	}
      \]
    }
  }

  The reduction is as follows:
  \[
    \begin{array}{r@{\ }l}
      \multicolumn{2}{l}{\pi_{((R\wedge S)\Rightarrow R)\wedge T}((\lambda x^{R\wedge S}.x)+\ve r)}\\
      &\eq
      \pi_{(R\wedge S)\Rightarrow R}(\lambda x^{R\wedge S}.x)+\pi_T(\ve r)\\
      &\re
      \pi_{(R\wedge S)\Rightarrow R}(\lambda x^{R\wedge S}.x)+\ve r\\
      &\eq
      (\lambda x^{R\wedge S}.\pi_R(x))+\ve r
    \end{array}
  \]
\end{example}

\subsection{Correctness}
The correctness of this calculus has been proved in~\cite{DiazcaroDowek15}.

The first result is the Subject Reduction property.

\begin{theorem}[Subject Reduction \citaTeoOrig{2.1}] 
  If $\Gamma\vdash\ve r:R$ and $\ve r\re\ve s$ or $\ve r\eq\ve s$, then $\Gamma\vdash\ve s:R$.
  \qed
\end{theorem}

The second result is the strong normalisation property. In our setting, strong normalisation means that every reduction sequence fired from a typed term eventually terminates in a term in {\em normal form} modulo $\eq^*$. In other words, 
no $\re$ reduction can be fired from it, even after $\eq$ steps.
Formally, 
we define $\Red{\ve r}=\{\ve s~|~\ve r\toreq\ve s\}$. Hence, a term $\ve r$ is in normal form if $\Red{\ve r}=\emptyset$. 

\begin{theorem}
  [Strong Normalisation \citaTeoOrig{3.17}]
  If $\Gamma\vdash\ve r:R$, then $\ve r$ is strongly normalising.
  \qed
\end{theorem}

\section{Implementing Isomorphic Types}\label{sec:types}
In order to implement isomorphic types, we consider conjunctions as multiset constructors. Hence, $(R\wedge S)\wedge T$ is implemented as the multiset $\sms{R,S,T}$, and this is the same implementation for $(T\wedge R)\wedge S$, or any other type isomorphic to this one using only isomorphisms (comm) and (assoc). Therefore, we also consider $\sms{R,\sms{S,T}}=\sms{R,S,T}$.

Hence, the grammar of types is the following:
\[
  R,S := \tau~|~R\Rightarrow S~|~\ms{R_i}in
\]

The remaining isomorphisms, namely (dist) and (curry), are implemented by its canonical form, as defined below.

\begin{definition}[Canonical form of a type]
  \label{def:canonical}
  The canonical form of a type is defined inductively by
  \[
    \begin{array}{r@{~=~}l@{}l}
      \can\tau & \tau &\\
      \can{R\Rightarrow S} & \text{let } & \ms{T_i\Rightarrow\tau}in= \can S\\
      \multicolumn{1}{c}{} &\text{in } & \ms{\can R\uplus\sms{T_i}\Rightarrow \tau}in\\
      \can{\ms{R_i}in} & \multicolumn{2}{l}{\biguplus_{i=1}^n\can{R_i}}\\
    \end{array}
  \]
where $\uplus$ is the union of multisets.
\end{definition}

Note that if $\can S$ is not shaped $\ms{T_i\Rightarrow\tau}in$, then $\can{R\Rightarrow S}$ is ill defined. Nevertheless, this never happens, as shown by the following lemma:
\begin{lemma}
  \label{lem:canShape}
  The canonical form of a type is produced by the following grammar:
  \(
    C := \ms{C_i\Rightarrow\tau}in
  \),
  with the following conventions:
    $\ms{C_i}i0\Rightarrow\tau = \tau$
    and
    $\ms{C_i}i1=C_1$.
\end{lemma}
Transforming a canonized type into a type from the original type system can be done just by grouping the types on a multiset choosing an arbitrary way to associate. Hence, we define the $\uncan{\cdot}$ function which does exactly that.
\begin{definition}
\(
  \uncan{\ms{C_i\Rightarrow\tau}in} = \bigwedge_{i=1}^n(\uncan{C_i}\Rightarrow\tau)
\)
where the big conjunction symbol associates to the left.
\end{definition}

From now on, we use $R$, $S$ and $T$ for generic types and $C$ and $D$ for canonical types.

The main idea introducing canonical types is that isomorphic types have the same canonical forms (Lemma~\ref{lem:uniqueNF}), and the canonical form is isomorphic to it (Lemma~\ref{lem:eqCan}).
\begin{lemma}
  \label{lem:uniqueNF}
  If $R\equiv S$, then $\can R=\can S$.
  \qed
\end{lemma}

\begin{lemma}
  \label{lem:eqCan}
  For any $R$, $R\equiv\can R$.
  \qed
\end{lemma}

Since types are part of the terms, we extend the definition of $\can{\cdot}$ to terms by taking the canonical form of its types. Also, we use multisets for sums in order to deal with the associativity and commutativity properties of this operator. Hence, the new grammar of terms is given by:
\[
  \ve r, \ve s::=\quad x^C~|~\lambda x^C.\ve r~|~\ve r\ve s~|~\ms{\ve r_i}in~|~\pi_C(\ve r)
\]
\begin{definition}[Canonical form of a term]
  \begin{align*}
    \can{x^R} &= x^{\can R}\\
    \can{\lambda x^R.\ve r} &= \lambda x^{\can R}.\canv r\\
    \canv{rs} &= \canv r\canv s\\
    \can{\ve r+\ve s} &= \sms{\canv r,\canv s}\\
    \can{\pi_R(\ve r)} &=\pi_{\can R}(\canv r)
  \end{align*}
\end{definition}

We also trivially extend $\can{\cdot}$ to contexts, by applying $\can{\cdot}$ to each typed variable of it.

A type system typing terms with canonical types is presented in Figure~\ref{fig:cantypes}.
The empty multiset is not a valid type, since the empty conjunction neither is in $\lambda^+$. However, we may write a type $[R]\uplus[S]$ where $[R]$ is empty and $[S]$ is not, since the whole type is not empty. Also, we may write $\gms{R}$ for $\ms{R_i}in$, when we are not interested in the number of types in the multiset, and $\gms{R}_k$ for different multisets parametrized by $k$. 

\begin{figure*}[!t]
  \centering
  \[
    \infer[^{(\textsl{ax})}]{\Gamma,x^{\gms C}\vdash x^{\gms C}:\gms C}{}
    \quad
    \infer[^{(\Rightarrow_i)}]{\Gamma\vdash\lambda x^{\gms C}.\ve r:\ms{\gms C\uplus \sms{D_i}\Rightarrow\tau}in}{\Gamma,x^{\gms C}\vdash\ve r:\ms{D_i\Rightarrow\tau}in}
    \quad
    \cond[\_]{\gms D\subseteq\bigcap\limits_{k=1}^m \gms C_k}
    \infer[^{(\Rightarrow_e)}]
    {
      \Gamma\vdash\ve r\ve s:
      \ms{
	\gms C_k\setminus\gms D\Rightarrow\tau
      }km
    }{
      \Gamma\vdash\ve r:\ms{\gms C_k\Rightarrow\tau}km
      &
    \Gamma\vdash\ve s:\gms D}
  \]
  \[
    \infer[^{(\wedge_i)}]{\Gamma\vdash\ms{\ve r_i}in:\ms{C_i}in}{(i=1,\dots,n)\ \Gamma\vdash\ve r_i:\sms{C_i}}
    \qquad
    \cond{\gms D\subseteq\gms C}\infer[^{(\wedge_e)}]{\Gamma\vdash\pi_{\gms D}\ve r:\gms D}{\Gamma\vdash\ve r:\gms C}
  \]
  \caption{Modified type system assigning only canonical types}
  \label{fig:cantypes}
\end{figure*}

Theorem~\ref{thm:TypeEq} shows that the type system given in Figure~\ref{fig:cantypes} is sound and complete with respect to the type system of $\lambda^+$ shown in Figure~\ref{fig:origTypes}. To this end, we define $\uncan{\ve r}$ which transforms terms written with multisets into terms written with sums, as follows.
\begin{definition}
  \begin{align*}
    \uncan{x^R} &= x^{\uncan R}\\
    \uncan{\lambda x^R.\ve r} &= \lambda x^{\uncan R}.\uncan{\ve r}\\
    \uncan{\ve r\ve s} &= \uncan{\ve r}\uncan{\ve s}\\
    \uncan{\ms{\ve r_i}in} &= \sum_{i=1}^n\uncan{\ve r_i}\\
    \uncan{\pi_R(\ve r)} &=\pi_{\uncan R}{(\uncan{\ve r})}
  \end{align*}
  where $\sum_{i=1}^n\ve r_i$ associates to the left.
\end{definition}

\begin{lemma}\label{lem:canuncan}
  $\can{\uncan{\ve r}}=\ve r$ and $\uncan{\can{\ve r}}\eq\ve r$.
  \qed
\end{lemma}

\begin{theorem}
  \label{thm:TypeEq}\conlista
  \begin{enumerate}
    \item 
      If $\Gamma\vdash\ve r:R$ is derivable in $\lambda^+$, then
      \[
	\can{\Gamma}\vdash\canv r:\can{R}
      \] is derivable in the modified system from Figure~\ref{fig:cantypes}.
    \item If $\Gamma\vdash\ve r:C$ is derivable in the modified system from Figure~\ref{fig:cantypes}, then \[
	\uncan\Gamma\vdash\uncan{\ve r}:\uncan C
      \] is derivable in $\lambda^+$.
      \qed
  \end{enumerate}
\end{theorem}

\section{Implementing the Rewrite Relation Modulo}\label{sec:rewrite}
In order to implement the rewrite relation modulo, we modify the rewrite system and get rid of the relation $\eq$. 
The modified rewrite system is presented in Figure~\ref{fig:modRewrite}, and 
Theorem~\ref{thm:rewSoundComp} shows its soundness.

Relation $\to$ is defined as $\stackrel{\neg\delta}{\longrightarrow}\cup\stackrel{\delta}{\longrightarrow}$.
The new rewrite rules can be understood reading them in order:
%The rule (uncurry) would be the first to be applied, if possible. This rule will transform lambda abstractions uncurryfying them. 
Rules ($\beta$) and (p$\beta$) are two ways to apply the $\beta$-reduction, the former when the type of the argument has the type expected by the abstraction, and the latter, ``partial-$\beta$'', when the argument is included between the list of expected arguments.
Finally, rule (d$\beta$), the ``delayed-$\beta$'' is meant to solve the application of a function to an argument of a different type. For example, consider the abstraction $\lambda x^{\sms{(\tau\Rightarrow\tau),\tau}\Rightarrow\tau}.x(\lambda y^\tau.y)$. The expected argument is a function taking $\tau\Rightarrow\tau$ and returning $\tau\Rightarrow\tau$, or, what is the same, a function expecting $\sms{(\tau\Rightarrow\tau),\tau}$ and returning $\tau$. Since in the body of the abstraction an argument of type $\tau\Rightarrow\tau$ is provided,
the type of the full term is 
\(
  (\sms{(\tau\Rightarrow\tau),\tau}\Rightarrow\tau)\Rightarrow(\tau\Rightarrow\tau)
\),
which, in canonical form, is
\(
  \sms{\sms{(\tau\Rightarrow\tau),\tau}\Rightarrow\tau,\tau}\Rightarrow\tau
\).
So, we can apply it to a term of type $\tau$. Let $\vdash\ve r:\tau$, then
\(
  (\lambda x^{\sms{(\tau\Rightarrow\tau),\tau}\Rightarrow\tau}.x(\lambda y^\tau.y))\ve r
\)
is typable. However, notice that the argument $\ve r$ must be delayed, because $\tau$ is not included in the type of $x$. Indeed, a function expecting an argument of type $\tau$ will be issued from $x(\lambda y^\tau.y)$. This case is when the rule (d$\beta$) applies, producing
\(
  \lambda x^{\sms{(\tau\Rightarrow\tau),\tau}\Rightarrow\tau}.(x(\lambda y^\tau.y)\ve r)
\).

Rule (curry) can be applied when all the $\beta$ rules have failed. It covers the case when part of the argument is included in the expected arguments, and another part needs to be delayed.

Rules (dist$_i$), (comm$_{ei}$) and (comm$_{ee}$) are direct consequences of rules \terrulelabel{(dist$_{ie}$)}, \terrulelabel{(dist$_{ei}$)} and \terrulelabel{(dist$_{ee}$)} of $\lambda^+$ respectively.

Rules (proj), (simp) and (dist$_e$) implement the projection (rules \terrulelabel{($\pi_n$)} and \terrulelabel{($\pi_1$)}) recursively: when the term projected has the type to be projected, the whole term is returned. This is the base case (rule (proj)). If, instead, the term projected is bigger, it is simplified (rule (simp)). Finally, when the type projected is a multiset of types, it can be split (rule (dist$_e$)).

The last rule is the surjective pairing, ($\delta$), which is meant to be applied only when nothing else applies, and the term context is appropriate. The contextual rules ($C_\to^{(\neg\delta)}$) and ($C_\to^\delta$) prevent ($\delta$) from being applied under a projection, and are direct consequences of the homonymous rules in $\lambda^+$.

\begin{figure*}[!t]
  \centering
  \noindent\begin{tabular}{rl}
%    (uncurry) &
%    $\lambda x^{\gms C}.\lambda y^{\gms D}.\ve r
%    \stackrel{\neg\delta}{\longrightarrow}
%    \lambda z^{\gms C\uplus\gms D}.\ve r\left\{\pi_{\gms C}(z)/x, \pi_{\gms D}(z)/y\right\}$
%    \\
    ($\beta$) &
    \(
    (\lambda x^{\gms C}.\ve r)\ve s~\stackrel{\neg\delta}{\longrightarrow}~\ve r\{\ve s/x\}
    \)
    \qquad
    where
    \qquad
    $\Gamma\vdash\ve s:\gms C$
    \\
    (p$\beta$) &
    \(
    (\lambda x^{\gms C}.\ve r)\ve s~\stackrel{\neg\delta}{\longrightarrow}~\lambda y^{\gms C\setminus\gms D}.\ve r\{\sms{\ve s,y}/x\}
    \)
    \qquad
    where 
    \qquad
    $\left\lbrace\begin{array}{l}
      \Gamma\vdash\ve s:\gms D\\
      \gms D\subset \gms C
    \end{array}\right.$
    \\
    (d$\beta$) &
    \(
    (\lambda x^{\gms C}.\ve r)\ve s~\stackrel{\neg\delta}{\longrightarrow}~\lambda x^{\gms C}.(\ve r\ve s)
    \)
    \qquad
    where
    \qquad
    $\left\lbrace\begin{array}{l}
      \Gamma\vdash\ve s:\gms D\\
      \gms D\cap\gms C=\emptyset
    \end{array}\right.$
    \\
    (curry) &
    $\ve r\ms{\ve s_i}in~\stackrel{\neg\delta}{\longrightarrow}~\ve r\ve s_1\dots\ve s_n$
    \\
    (dist$_i$) & $\ms{\ve r_i}in\ve s\stackrel{\neg\delta}{\longrightarrow}\ms{\ve r_i\ve s}in$
    \\
    (comm$_{ei}$) & $\pi_{\ms{\gms C_k\Rightarrow\tau}km}(\lambda x^{\gms D}.\ve r)
    ~\stackrel{\neg\delta}{\longrightarrow}~
    \lambda x^{\gms D}.\pi_{\ms{(\gms C_k\setminus\gms D)\Rightarrow\tau}km}(\ve r)$
    \hspace{1mm} where
    $\gms D\subseteq \bigcap\limits_{k=1}^m \gms C_k$
    \\
    (comm$_{ee}$) & $\pi_{\ms{C_i\Rightarrow\tau}in}(\ve r\ve s)\stackrel{\neg\delta}{\longrightarrow}\pi_{\ms{\gms D\uplus\sms{C_i}\Rightarrow\tau}in}(\ve r)\ve s$
    \qquad
    where
    \qquad
    $\Gamma\vdash\ve s:\gms D$.
    \\
    (proj) &
    $\pi_{\gms C}(\ve r)~\stackrel{\neg\delta}{\longrightarrow}~\ve r$
    \qquad
    where
    \qquad
    $\Gamma\vdash\ve r:\gms C$.
    \\
    (simp) &
    $\pi_{\gms C}(\gms{\ve r}\uplus\gms{\ve s})~\stackrel{\neg\delta}{\longrightarrow}~\pi_{\gms C}\gms{\ve r}$
    \qquad
    where
    \qquad
    $\left\{\begin{array}{l}
      \Gamma\vdash\gms{\ve r}:\gms D\\
      \gms C\subseteq\gms D
    \end{array}\right.$
    \\
    (dist$_e$) &
    \(
    \pi_{\gms C\uplus\gms D}(\gms{\ve r}\uplus\gms{\ve s})~\stackrel{\neg\delta}{\longrightarrow}~\sms{\pi_{\gms C}\gms{\ve r},\pi_{\gms D}\gms{\ve s}}
    \)
    \qquad
    where
    \qquad
    $\left\{\begin{array}{l}
      \Gamma\vdash\ve r:\gms{C'}\\
      \Gamma\vdash\ve s:\gms{D'}\\
      \gms{C'}\subseteq\gms C\\
      \gms{D'}\subseteq\gms D
    \end{array}\right.$
    \\
    ($\delta$) &
    \(
    \ve r~\stackrel{\delta}{\longrightarrow}~\ms{\pi_{\sms{C_i}}(\ve r)}in
    \)
    \ 
    where
    $\left\lbrace\begin{array}{l}
      \Gamma\vdash\ve r\neq\ms{\ve r_i}in,
      \textrm{ with }\Gamma\vdash\ve r_i:\sms{C_i}
      \\
      \Gamma\vdash\ve r:\ms{C_i}in
    \end{array}\right.$
  \end{tabular}

  \[
    \infer[^{(C_{\to}^{(\neg\delta)})}]{C[\ve r]\stackrel{\neg\delta}{\longrightarrow} C[\ve s]}{\ve r\stackrel{\neg\delta}{\longrightarrow}\ve s}
    \qquad
    \infer[^{(C_{\to}^{\delta})}]{C[\ve r]\stackrel{\delta}{\longrightarrow}C{[\ve s]}}
    {
      \ve r\stackrel{\delta}{\longrightarrow}\ve s
      &
      C[\cdot]\neq C'[\pi_R(\cdot)]
    }
  \]
  \caption{Rewrite system of the implementation}
  \label{fig:modRewrite}
\end{figure*}

Theorem~\ref{thm:rewSoundComp} shows the soundness of relation $\to$ with respect to the original relations $\eq$ and $\re$. In particular, we decided to treat functions and curryfication in a different way, by rules (p$\beta$), (d$\beta$) and (curry), and hence if a function is equivalent to another term in $\lambda^+$, it may not happen the same in our implementation, however, functions behave the same in the sense that when they are applied to an argument, they produce the same result.

\begin{theorem}
  \label{thm:rewSoundComp}
  Let $\ve r$ be a closed term in $\lambda^+$.
  \begin{itemize}
    \item If $\ve r\eq\ve r'$ by any rule other than AC, then 
      \begin{itemize}
	\item 
	  if $\vdash\ve r:R\Rightarrow S$, then for all $\vdash\ve s:\can R$, where $\ve s$ is in canonical form, there exists $\ve t$ such that
	  \begin{center}
	    \begin{tikzpicture}
	      \node (u) at (0,0) {$\can{\ve r}\ve s$};
	      \node (d) at (2,0) {$\can{\ve r'}\ve s$};
	      \node (t) at (1,-1) {$\ve t$};	    
	      \path (u) edge[to*] (t) (d) edge[to*] (t);
	    \end{tikzpicture}
	  \end{center}
	\item Otherwise, there exists $\ve t$ such that 
	  \begin{center}
	    \begin{tikzpicture}
	      \node (u) at (0,0) {$\can{\ve r}$};
	      \node (d) at (2,0) {$\can{\ve r'}$};
	      \node (t) at (1,-1) {$\ve t$};	    
	      \path (u) edge[to*] (t) (d) edge[to*] (t);
	    \end{tikzpicture}
	  \end{center}
      \end{itemize}
    \item If $\ve r\hookrightarrow\ve r'$, then $\can{\ve r}\to^+\can{\ve r'}$.
      \qed
  \end{itemize}
\end{theorem}

\section{$N$-tuples, Natural Numbers and Recursion}\label{sec:nat}
\subsection{Justification}
Notice that in the terms $\lambda x^\tau.\lambda y^\tau.x$ and $\lambda x^\tau.\lambda y^\tau.y$ it cannot be ensured which argument will be returned. Indeed, only the order given to its arguments in the implementation will choose which argument to return. Hence, the classical Church encodings cannot work\footnote{Of course, we would need polymorphism for Church encodings. However, even if we extend this calculus with polymorphism, Church encodings will not work.}.
Therefore, we extend our calculus with primitive natural numbers. In addition, we include general recursion to increase the expressiveness of the resulting language.

In any case, natural numbers and recursion are not enough. Consider the subtraction of two natural numbers
\[
  \s{subtraction} = \lambda x^{Nat}.\lambda y^{Nat}.x-y
\]
Again, which argument is evaluated first depends on the specific implementation. 
Hence, $\s{subtraction}\ 3\ 2$ may reduce in the same way as $\s{substraction}\ 2\ 3$.

To solve these problems we need to distinguish arguments of the same type. In particular, we need a tuple so $\s{subtraction}$ takes the tuple and calculates the first minus the second. This can be done with an encoding.

In Section~\ref{sec:encoding} we present an encoding for deterministic $n$-tuples by extending the calculus with a second atomic type $\iota$. In Section~\ref{sec:natural} we extend the resulting calculus with natural numbers and structural recursion.

\subsection{$N$-tuples}\label{sec:encoding}
In this section we add an encoding for deterministic $n$-tuples, which will be handy to use together with natural numbers, introduced in Section~\ref{subsec:nat}. To this end, we need to distinguish the types of the first element of the tuple, the second element of the tuple, and so on. Thus, we can project an element with respect to its ``position''. We will use a new type constant for this, $\iota$, so we can be sure that it is not used somewhere else.

Consider the following encoding:
\[
  \num 1 = \sms\iota,\quad
  \num 2 = \sms{\iota,\iota}\quad
  \num 3 = \sms{\iota,\iota,\iota},\quad
  \dots\quad
  \num n = \sms{\iota,\dots,\iota}
\]

Let $\bnum n=\num n\Rightarrow\iota$. The $n$-tuple $(\ve r_1,\dots,\ve r_n)$ is defined by
\[
  (\ve r_1,\ve r_2,\dots,\ve r_n) := \ms{\lambda x^{\bnum i}.\ve r_i}in
\]
where for all $k$, $\Gamma\vdash\ve r_k:C_k$, $x^{\bnum i}\notin\Gamma$ and $C_k\not\equiv\bnum j\Rightarrow D$, for any $\bnum j\in\{\bnum 1,\dots,\bnum n\}$.

Hence, every term in the multiset has a different type, so we can project with respect to such a type to obtain the element. After projecting the $i$-th component, it has to be applied to a term of type $\bnum i$ to recover the original term. Let $\estr n=\lambda x^{\num n}.\pi_\iota x$. Hence, the tuples projectors are defined by
\begin{align*}
  \fstO{C_1}(\ve r) &:= (\pi_{\can{\bnum 1\Rightarrow C_1}}(\ve r))\estr 1\\
  \sndO{C_2}(\ve r) &:= (\pi_{\can{\bnum 2\Rightarrow C_2}}(\ve r))\estr 2\\
  &\vdots\\
  \s{nth}_{C_n}(\ve r) &:= (\pi_{\can{\bnum n\Rightarrow C_n}}(\ve r))\estr n
\end{align*}

Notice that because of the restriction of $C_k\not\equiv\bnum j\Rightarrow D$, the encoding cannot be defined generically. For example,
let $\Gamma\vdash\ve r:\bnum 2\Rightarrow\tau$ and $\Gamma\vdash\ve s:\bnum 1\Rightarrow\tau$, then
\[
  \fstO{{\bnum 2\Rightarrow\tau}}(\ve r,\ve s) 
  :=(\pi_{{\sms{\bnum 1,\bnum 2}\Rightarrow\tau}}\sms{\lambda x^{\bnum 1}.\ve r,\lambda x^{\bnum 2}.\ve s})\estr 1
\]
However, this projection is non-deterministic. Indeed,
\[
  \Gamma\vdash\lambda x^{\bnum 1}.\ve r:\sms{\bnum 1,\bnum 2}\Rightarrow\tau\,,
  \quad\text{but also}\quad
  \Gamma\vdash\lambda x^{\bnum 2}.\ve s:\sms{\bnum 1,\bnum 2}\Rightarrow\tau
\]

A workaround is not to use the numbers $\bnum 1$ or $\bnum 2$ for the encoding, for example:
\[
  \fstO{\bnum 2\Rightarrow\tau}(\ve r,\ve s):=(\pi_{\sms{\bnum 3,\bnum 2}\Rightarrow\tau}\sms{\lambda x^{\bnum 3}.\ve r,\lambda x^{\bnum 4}.\ve s})\estr 3
\]
which means that it will be necessary to choose the right encoding for each tuple according to the type of its elements.

We use $\canon{\ve r}n$ for $\lambda w^{\bnum n}.\ve r$, where $w\notin FV(\ve r)$. Notice that $\canon{\ve r}n\estr n\to\ve r$. We also extend this notation to types, so $\canon{C}n = \can{\bnum n\Rightarrow C}$. In addition, for the sake of simplifying the notation, we may use $C\times D$ for $\can{\sms{\canon C1,\canon D2}}$.
Finally, we sometimes omit the canonicity function. For example, we may write $\sms C\Rightarrow\sms D$ for $\can{\sms C\Rightarrow\sms D}$. In general, any non-canonical type $R$ is just a notation for $\can R$.

\subsection{Natural Numbers and General Recursion}\label{sec:natural}
\label{subsec:nat}
In this section we add natural numbers and recursion.
We include $0$, the successor and predecessor, and for convenience a test for $0$ and a test for equality.
The new grammar of terms and types is the following:
\begin{align*}
  \ve r,\ve s, \ve z, \ve n, \ve m &:= x^C~|~\lambda x^C.\ve r~|~\ve r\ve s~|~\ms{\ve r_i}in~|~\pi_C(\ve r)\\
  &\quad |~0~|~\suc{\ve n}~|~\pred{\ve n}\\
  &\quad |~\ifZ~\ve n\ve r\ve s~|~\ifEq~\ve n\ve m\ve r\ve s\\
  &\quad |~\mu x^C.\ve r\\
  \tau &:=\iota~|~Nat\\
  C &:=\ms{C_i\Rightarrow\tau}in
\end{align*}

The type system from Figure~\ref{fig:cantypes} is extended with the rules from Figure~\ref{fig:nattypes}.
\begin{figure*}[!t]
  \centering
  \[
    \infer[^{(\textsl{ax}_0)}]{\Gamma\vdash 0:Nat}{}
    \quad
    \infer[^{(\textsl{succ})}]{\Gamma\vdash (\suc{\ve r}):Nat}{\Gamma\vdash\ve r:Nat}
    \quad
    \infer[^{(\textsl{pred})}]{\Gamma\vdash (\pred{\ve r}):Nat}{\Gamma\vdash\ve r:Nat}
  \]
  \[
    \infer[^{(\textsl{ifZ})}]{\Gamma\vdash \ifZ~\ve n\ve r\ve s:\gms C}
    {
      \Gamma\vdash\ve n:Nat
      &
      \Gamma\vdash\ve r:\gms C
      &
      \Gamma\vdash\ve s:\gms C
    }
    \qquad
    \infer[^{(\textsl{ifEq})}]{\Gamma\vdash \ifEq~\ve n\ve m\ve r\ve s:\gms C}
    {
      \Gamma\vdash\ve n:Nat
      &
      \Gamma\vdash\ve m:Nat
      &
      \Gamma\vdash\ve r:\gms C
      &
      \Gamma\vdash\ve s:\gms C
    }
  \]
  \[
    \infer[^{(\textsl{fix})}]{\Gamma,\vdash \mu x^{\gms C}.\ve r:\gms C}
    {\Gamma,x^{\gms C}\vdash\ve r:\gms C}
  \]
  \caption{Added typing rules for natural numbers and general recursion}
  \label{fig:nattypes}
\end{figure*}

The rewrite system from Figure~\ref{fig:modRewrite} is extended with the rules from Figure~\ref{fig:natrewrite}. Rules $\stackrel{\neg\delta}{\longrightarrow}$ are now notated $\stackrel{\neg\delta,\neg\mu}{\longrightarrow}$ and $\to=\stackrel{\neg\delta,\neg\mu}{\longrightarrow}\cup\stackrel{\delta}{\longrightarrow}\cup\stackrel{\mu}{\longrightarrow}$.
\begin{figure*}[!t]
  \centering
  \noindent\begin{tabular}{rl}
    (pred) &
    $\pred{(\suc{\ve n})}\stackrel{\neg\delta,\neg\mu}{\longrightarrow}\ve n$
    \\
    (ifZ$_0$) &
    $\ifZ~0\ve r\ve s \stackrel{\neg\delta,\neg\mu}{\longrightarrow} \ve r$
    \\
    (ifZ$_n$) &
    $\ifZ~(\suc{\ve n})\ve r\ve s \stackrel{\neg\delta,\neg\mu}{\longrightarrow} \ve s$
    \\
    (ifEq$_0$) &
    $\ifEq~0\ve m\ve r\ve s \stackrel{\neg\delta,\neg\mu}{\longrightarrow} \ifZ~\ve m\ve r\ve s$
    \\
    (ifEq$_n$) &
    $\ifEq~(\suc{\ve n})\ve m\ve r\ve s\stackrel{\neg\delta,\neg\mu}{\longrightarrow}\ifZ~\ve m\ve s(\ifEq~\ve n(\pred{\ve m})\ve r\ve s)$
    \\
    ($\mu$) &
    $\mu x^{\gms C}.\ve r \stackrel{\mu}{\longrightarrow} \ve r\repl{\mu x^{\gms C}.\ve r/x}$
    \\
    (comm$_{\textsl{ifZ}}$) &
    $\pi_{\gms C}(\ifZ~\ve n\ve r\ve s) \stackrel{\neg\delta,\neg\mu}{\longrightarrow} \ifZ~\ve n(\pi_{\gms C}\ve r)(\pi_{\gms C}\ve s)$
    \\
    (comm$_{\textsl{ifEq}}$) &
    $\pi_{\gms C}(\ifEq~\ve n\ve m\ve r\ve s)\stackrel{\neg\delta,\neg\mu}{\longrightarrow}\ifEq~\ve n\ve m(\pi_{\gms C}\ve r)(\pi_{\gms C}\ve s)$
    \\[1ex]
    (comm$_{\mu}$) &
    $\pi_{\gms{C_1}}(\mu x^{\gms{C_1}\uplus\gms{C_2}}.\ve r)
    \stackrel{\neg\delta,\neg\mu}{\longrightarrow}
    \mu x_1^{\gms{C_1}}.\pi_{\gms{C_1}}\ve r\repl
    {
      \sms{x_1,\mu x_2^{\gms{C_2}}.\pi_{\gms{C_2}}\ve r\repl{\sms{x_1,x_2}/x}}
    /x}$
  \end{tabular}
  \caption{Added rewrite rules for natural numbers and structural recursion}
  \label{fig:natrewrite}
\end{figure*}
Rule (comm$_{\mu}$) follows from Beki\'c's theorem~\cite{BekicLNCS}, which allows splitting a mutual recursion into two recursions, hence, in some cases, to simplify one of them (see examples in Section~\ref{sec:example}).
The contextual rules have to be updated too. The updated rules are depicted in Figure~\ref{fig:updatedContext}. 
\begin{figure}
  \[
    \begin{array}{c}
    \infer[^{(C_{\to}^{(\neg\delta,\neg\mu)})}]{C[\ve r]\stackrel{\neg\delta,\neg\mu}{\longrightarrow} C[\ve s]}{\ve r\stackrel{\neg\delta,\neg\mu}{\longrightarrow}\ve s}
    \\[1em]
    \infer[^{(C_{\to}^{\delta})}]{C[\ve r]\stackrel{\delta}{\longrightarrow}C{[\ve s]}}
    {
      \ve r\stackrel{\delta}{\longrightarrow}\ve s
      &
      C[\cdot]\neq C'[\pi_R(\cdot)]
    }
    \\[1em]
    \infer[^{(C_{\to}^{\mu})}]{C[\ve r]\stackrel{\mu}{\longrightarrow}C{[\ve s]}}
    {
      \ve r\stackrel{\mu}{\longrightarrow}\ve s
      &
      C[\cdot]\neq C'[\lambda x^{\gms D}.C''[\cdot]]
    }
    \\[1em]
    \infer[^{C_{[]}^\mu}]{\canon{\ve r}n\stackrel\mu\longrightarrow\canon{\ve s}n}
    {\ve r\stackrel\mu\longrightarrow\ve s}
  \end{array}
  \]
  \caption{Updated contextual rules}
  \label{fig:updatedContext}
\end{figure}
The new rule $(C_\to^\mu)$ prevents looping of the $(\mu)$ rule. In general, a way to prevent $(\mu)$ from looping by applying only this rule \emph{ad infinitum} is to forbid reduction under $\lambda$. In this system, we want reduction under $\lambda$ since we want to optimize functions. Hence, we only disallow reduction under $\lambda$ for the $\mu$ rule. Rule $(C_{[]}^\mu)$, on the other hand, allows to reduce $\mu$ under encodings. Notice that encodings are written with abstractions, however, the type in the argument of the abstraction distinguishes them from normal abstractions.

\section{Examples of Projecting Recursive Functions}\label{sec:example}
\subsection{Discarding Code}
In this section we present an example of projecting a recursive function which gets rid of unnecessary code. We define the function $\divMod{}{}$ as the function taking two natural numbers and returning the result of the integer division of the first number by the second number, together with the remainder of such a division.
We then define the integer division as the first projection over $\divMod{}{}$. The novelty is that such a projection will enter in the recursion to simplify the code by discarding the part of the code calculating the remainder.

First we need some auxiliary definitions. For reference, similar definitions in Haskell are detailed in Figure~\ref{fig:haskell}. 

\begin{figure}[!t]
\begin{verbatim}
succFst (x,y) = (x+1,y)

divModRec n m k =
  if (n==0)
   then (0,k)
   else if m == k+1
         then succFst (divModRec (n-1) m 0)
         else divModRec (n-1) m (k+1)

divMod (n,m) = divModRec n m 0

div (n,m) = fst (divMod (n,m))
   -- Notice that in Haskell we cannot apply
   -- fst to the function but to the result.
\end{verbatim}
  \caption{Definition of \s{div} in Haskell}
  \label{fig:haskell}
\end{figure}

%The function $\ifEq{}{}Z$ is an if-then-else construction where the condition is the equality of two natural numbers.
%The first natural number is taken as a term of type $Nat$, instead, the second will be encoded into $\cnat i$, where $\num i$ is a parameter depending on which is the first encoding for tuples that is not used. In the same way, the conditions are distinguished by marking one with $\cnat j$:
%\begin{align*}
%  \ifEq ij_{\gms C} := &
%  \mu x^{\sms{Nat,\cnat i,\gms C,\canon{\gms C}j}\Rightarrow\gms C}.\\
%  &\ \lambda n^{Nat}.\lambda m^{\cnat i}.\lambda r^{\canon{\gms C}j}.\lambda s^{\gms C}.\\
%  &\quad\ifZ~n\\
%  &\qquad(\ifZ~(m\estr i)(r\estr j)s)\\
%  &\qquad(\ifZ~(m\estr i)s(x(\pred n)(\pred m)rs))
%\end{align*}

The function $\succFst$ takes a tuple of two $Nat$ and increments by one the first one.
\[
  \succFst := \lambda x^{Nat\times Nat}. (\suc(\fst(x)), \snd(x))
\]

The function $\divModRec{}{}$ receives two natural numbers and returns a function that, given an accumulator counting the current remainder of the division so far, completes the task of division. The property that \divModRec{}{} satisfies is that $\divModRec{}{}\ \ve n \ve m \ve k = \divMod{}{}(\ve n+\ve k) \ve m$.
The accumulator is taken as a term of type $Nat$, instead, the two numbers will be encoded into $\cnat i$ and $\cnat j$, where $\num i$ and $\num j$ depend on which are the non-used encoding for lists.
\begin{align*}
  &\divModRec ij:=\\
  &\ \mu x^{\sms{\cnat i,\cnat j,Nat}\Rightarrow Nat\times Nat}.\\
  &\quad\lambda n^{\cnat i}.\lambda m^{\cnat j}.\lambda k^{Nat}.\\
  &\qquad\quad\begin{aligned}[t]
    \ifZ~&(n\estr i)~(0,k)\\
	 &\begin{aligned}[t]
	   (\ifEq~&(m\estr j)~(\suc k)\\
	   &(\succFst(x\canon{\pred{(n\estr i)}}im0))\\
	   &(x\canon{\pred{(n\estr i)}}im(\suc k)))
	  \end{aligned}
	\end{aligned}
\end{align*}

The function $\divMod{}{}$ just receives the arguments in a tuple, pass them to \divModRec{}{} in the right order, and initialises the accumulator to $0$.
\begin{align*}
  &\divMod ij :=\\
  &\quad\lambda x^{Nat\times Nat}.\divModRec ij\canon{\fst(x)}i\canon{\snd(x)}j 0
\end{align*}

Finally, we define $\s{div}$ as the first projection with respect to $\divMod{}{}$. 
Observe that we use the encodings 3 and 4 in this function, because the encodings 1 and 2
are used by the tuple of $Nat$.
\[
  \s{div} := \fstO{Nat\times Nat\Rightarrow Nat}(\divMod 34)
\]

The originality is that we can start reducing $\s{div}$ before the arguments arrive, which will optimize the code by erasing the non-used parts of the mutual recursion. This is a consequence of Beki\'c's theorem (rule (comm$_{\mu}$)) as well as the commutation rules (comm$_{ei}$) and (comm$_{ee}$) issued by the isomorphisms, which allows the projection to enter to the right place where it can act.

Hence, \s{div} reduces as shown in Figure~\ref{fig:divReduction}. The detailed trace of this reduction can be found in Appendix~\ref{ap:trace}.

\begin{figure}[!t]
  \centering
\begin{tabular}{*{50}{p{2mm}@{}}}
  \all{\s{div}\to^*}\\
  &\m{49}{(\lambda x^{Nat\times Nat}.}\\
  &&\m{48}{(\mu x_1^{\sms{\cnat 3,\cnat 4,Nat}\Rightarrow\cnat 1}.}\\
  &&&\m{47}{\lambda n^{\cnat 3}.\lambda m^{\cnat 4}.\lambda k^{Nat}.}\\
  &&&&\m{46}{\ifZ~(n\estr 3)}\\
  &&&&&\m{45}{\canon 01}\\
  &&&&&\m{45}{({\ifEq}~(m\estr 4)~(\suc k)}\\
  &&&&&&\m{44}{\canon{\suc((x_1\canon{\pred{(n\estr 3)}}3m0)\estr 1)}1}\\
  &&&&&&\m{44}{(x_1\canon{\pred{(n\estr 3)}}3m(\suc k))))}\\
  &&\m{48}{\canon{\fst(x)}3}\\
  &&\m{48}{\canon{\snd(x)}4}\\
  &&\m{48}{0}\\
  &\m{49}{)\estr 1}\\
\end{tabular}
  \caption{Reduction of \s{div}}
  \label{fig:divReduction}
\end{figure}

Notice that all the recursions on $Nat\times Nat$ became recursions on $Nat$.
For reference, a definition similar to the optimised code of \s{div} is shown in Figure~\ref{fig:redDiv}.

\begin{figure}[!t]
\begin{verbatim}
div (n,m) = divRec n m 0

divRec n m k =
    if (n==0)
     then 0
     else if m==k+1
           then divRec (n-1) m 0 +1
           else divRec (n-1) m (k+1)
\end{verbatim}
  \caption{Optimized \s{div} in Haskell}
  \label{fig:redDiv}
\end{figure}

\subsection{Mutual Recursion}
The previous example was somehow ``easy'': \divMod{}{} makes two calculations independent of each other, so \s{div} discards the pieces of code not needed for the calculation of the division alone. In this section we propose a second example where the mutual recursion is used in the calculation and so the code of one recursion is used in the second recursion.

Let \evenOdd\ be a function which, given a natural number, determines whether it is even or odd. The output of \evenOdd\ is a tuple of two elements, the first element being $0$ if the number is even or $\suc 0$ if it is not even, and the second returning $0$ if it is odd or $\suc 0$ if it is not odd. Such a function can be programmed by:
\begin{align*}
  \evenOdd &:= \mu x^{Nat\Rightarrow Nat\times Nat}.\lambda n^{Nat}.\\
  &\qquad\ifZ~n~(0,\suc 0)~(\swap~(x~(\pred~n)))
\end{align*}
where 
\(
  \swap := \lambda x^{Nat\times Nat}.(\snd x,\fst x)
\).

As a side note, remark that the \swap\ function takes a term of type 
\[
  Nat\times Nat=\sms{\cnat 1,\cnat 2}
\]
that is, an encoded pair. Then, \swap\ changes the encoding so the first element is encoded as the second and the second as the first:
\begin{align*}
  \swap &= \lambda x^{Nat\times Nat}.(\snd x,\fst x)\\
  &=\lambda x^{Nat\times Nat}.((\pi_{\cnat 2} x)\estr 2,(\pi_{\cnat 1} x)\estr 1)\\
  &=\lambda x^{\sms{\cnat 1,\cnat 2}}.\sms{\canon{(\pi_{\cnat 2} x)\estr 2}1,\canon{(\pi_{\cnat 1} x)\estr 1}2}
\end{align*}

Coming back to the \evenOdd\ function, we can check whether a number is even by projecting the first element of the output of \evenOdd\ and passing it to \ifZ\ in order to detect whether it is $0$ (even) or not (odd).

Hence, we can define:
\[
  \s{even} := \fstO{Nat\Rightarrow Nat}(\evenOdd)
\]
\begin{figure}[!t]
\begin{verbatim}
swap (x,y) = (y,x)

evenOdd n = if (n==0) 
             then (0,1)
             else swap (evenOdd (n-1))

even n = fst (evenOdd n)
\end{verbatim}
  \caption{Definition of \s{even} in Haskell}
  \label{fig:haskellEven}
\end{figure}
Similar definitions to \evenOdd\ and \s{even}, written in Haskell, can be found in Figure~\ref{fig:haskellEven} for easier reading.

Observe that the call to the ``odd'' function is expanded and only the important parts remain -- the rest is simplified by the rules. The reduction of \s{even} is shown in Figure~\ref{fig:evenReduction}. 
The detailed trace of this reduction can be found in Appendix~\ref{ap:traceEven}.

Notice that the argument $x_2$ from the second recursion is not used anywhere, however it is not simplified since it is under lambda. We could, nevertheless, add a rule for this kind of trivial $\mu$, where the argument is lost after one iteration, in the following way:
\begin{center}
  \begin{tabular}{rl}
    (t$\mu$) & $\mu x^C.\ve r \stackrel{\neg\delta,\neg\mu}{\longrightarrow}\ve r$\qquad if $x\notin FV(\ve r)$
  \end{tabular}
\end{center}
The reduction of \s{even} with rule (t$\mu$) is shown in Figure~\ref{fig:betterEvenReduction}.
For reference, a similar definition to the developed code of \s{even} with (t$\mu$) is shown in Figure~\ref{fig:redEven}.

\begin{figure}[!t]
  \centering
  \begin{tabular}{*{50}{p{2mm}@{}}}
    \all{\s{even}\to^*}\\
    &\m{49}{(\mu x_1^{Nat\Rightarrow\cnat 1}.\lambda n^{Nat}.}\\
    &&\m{48}{(\ifZ~n~\canon 01}\\
    &&&\m{47}{{\lbrack}({\mu x_2^{Nat\Rightarrow\cnat 2}.}}\\
    &&&&\m{46}{{(\ifZ~(\pred~n)}}\\
    &&&&&\m{45}{{\canon{\suc 0}2}}\\
    &&&&&\m{45}{{\canon{(x_1(\pred~(\pred~n))\estr 1)}2)}}\\
    &&&\m{47}{)\estr 2\rbrack^{\bnum 1})}\\
    &\m{49}{)\estr 1}
  \end{tabular}
  \caption{Reduction of \s{even}}
  \label{fig:evenReduction}
\end{figure}

\begin{figure}[!t]
  \centering
  \begin{tabular}{*{50}{p{2mm}@{}}}
    \all{\s{even}\to^*}\\
    &\m{49}{(\mu x_1^{Nat\Rightarrow\cnat 1}.\lambda n^{Nat}.}\\
    &&\m{48}{(\ifZ~n~\canon 01}\\
    &&&\m{47}{\lbrack(\ifZ~(\pred~n)}\\
    &&&&\m{46}{\canon{\suc 0}2}\\
    &&&&\m{46}{\canon{(x_1(\pred~(\pred~n))\estr 1)}2)}\\
    &&&\m{47}{\estr 2\rbrack^{\bnum 1})}\\
    &\m{49}{)\estr 1}
  \end{tabular}
  \caption{Reduction of \s{even} with (t$\mu$)}
  \label{fig:betterEvenReduction}
\end{figure}

\begin{figure}[!t]
  \centering
\begin{verbatim}
even n = if (n==0)
          then 0
          else if (n-1==0)
                then 1
                else even (n-2)
\end{verbatim}
  \caption{Developed definition of \s{even} in Haskell}
  \label{fig:redEven}
\end{figure}

\section{Conclusions and Future Work}\label{sec:conclusion}
In this paper we have proposed a non trivial implementation of $\lambda^+$~\cite{DiazcaroDowek15}. The main difficulty is the fact that $\lambda^+$ has a rewrite system modulo an equivalence relation. We proposed a modified type system where all the isomorphisms are solved by picking a representing member of its equivalence class (a canonical form). Also, we proposed a modified rewrite system where all the possibilities from the equivalence relation in the original rewrite system are directed. Finally, we extended the calculus with natural numbers and general recursion.

The obtained system has an enhanced form of partial application where a function can receive its arguments in any order (Rule p$\beta$) and even any subset of its arguments. Even more, a function returning a second function, can receive the arguments for the second function before the arguments for the function in the head position (Rule d$\beta$).

In addition, our system provides the ability of defining a function with mutual recursion and later projecting one of its results to simplify the function, thanks to the commutativity rules witch enter the projection deep enough to place the projection where it can act.

The prototype uses an algorithm to reconstruct types from the type decorations in the terms. Note that it is enough with that -- a type inference algorithm is not needed because the system is simply typed in Church style.

As mentioned earlier, the starting point of the language presented in this paper is the calculus $\lambda^+$ developed in~\cite{DiazcaroDowek15}. In $\lambda^+$ the aim was to develop a calculus closer to the logical interpretation of predicates: For example, the predicate $A\wedge B$ is not distinguished from the predicate $B\wedge A$, and so, the aim was to not distinguish their proofs neither.
Therefore, the logic behind our language is nothing more than first order logics with conjunction and implication. No modifications have been done at the logical level.
\medskip

\subsection{Related Works}
\paragraph{Selective $\lambda$-Calculus}
One may ask whether $\lambda^+$ is somhow minimal, or is there a calculus which does not have pairs, but also allows for non-determinism due to type isomorphism. 
Indeed, in a work by Garrigue and A\"it-Kaci~\cite{GarrigueAitkaciPOPL94}, only the isomorphism $(R\Rightarrow S\Rightarrow T) \equiv (S\Rightarrow R\Rightarrow T)$ has been treated, which is complete with respect to the function type. Notice that this isomorphism is also valid in $\lambda^+$, as a consequence of
$R\wedge S\equiv S\wedge R$ and 
$(R\wedge S)\Rightarrow T\equiv (R\Rightarrow S\Rightarrow T)$. 
Their proposal is the selective $\lambda$-calculus, a calculus including labellings to identify which argument is being used at each time. Moreover, by considering the Church encoding of pairs, this isomorphism implies the commutativity on pairs.
However their proposal is different to ours. 
In particular, we track the term by its type, which is a kind of labelling, but when two terms have the same type, then we leave the system to non-deterministically choose any of them. 
One of the main novelties of $\lambda^+$ is, indeed, the non-deterministic projector. In addition, the (dist) and (curry) isomorphisms are those giving the improvements shown in Section~\ref{sec:example}.

\paragraph{Intersection Types and Semantic Subtyping}
No direct connection seems to exists between $\lambda^+$ and intersection types~\cite{PimentelRonchiRoversiFI12,BarbaneraDezaniDeLiguoroIC95,DunfieldJFP14}. However, there is an ongoing project of a new type system based on intersections, which may take some of the ideas from \cite{DiazcaroDowek16}, which is a simplification of $\lambda^+$, with extensions for quantum computation.
In~\cite{DebenadettieRonchiITRS12} a type system with non-idempotent intersection has been used to compute a bound on the normalisation time, and in~\cite{BernadetLengrandLMCS13,KesnerVenturaLNCS14} to provide new characterisations on strongly normalising terms. In \cite{CastagnaNguyXuImLengletPadovaniPOPL14} the authors introduce a calculus with intersection types, showing a practical use in nowadays programming languages.
In $\lambda^+$, however, the sum, which resebles an intersection, is not used as a way of polymorphism or to give quantitative information. Instead our sum is really a multiset of terms (not only on types). Moreover, we focus more in the two isomorphisms (dist) and (curry), which provides most of the interesting features in $\lambda^+$, and not much on the pair/intersection construction. 

\subsection{Future Work}

There are several possible future directions that we are willing to pursue:

\paragraph{Defining and Studying Evaluation Orders}
We have not given any reduction strategy to reduce expressions. In order to have a useful execution mechanism, reduction orders have to be defined and studied. For example, applicative or normal orders can be considered and extended to take into account the new rules (mostly the commutative ones).

\paragraph{Removing Non-Determinism}
A natural direction to follow is to get rid of the non-determinism by removing isomorphisms (comm) and (assoc). The original interest of $\lambda^+$ has been to study non-determinism among other things, however it seems that isomorphisms (curry) and (dist) are all what we need in order to have the strong partial application and the ability to project functions presented in this paper. 

\paragraph{Studying Non-Determinism}
Another (opposite) direction is to study the non-determinism issued from this calculus. For example, the types encodings $\bnum n$ are inhabited by functions with different ``degree'' of non-determinism, as explained next.

Since $\iota$ is not inhabited, the only inhabitant of $\bnum 1 = \iota\Rightarrow\iota$ is $\lambda x^\iota.x$, or, what is equivalent $\lambda x^\iota.\pi_\iota x=\estr 1$.
Analogously, the only inhabitant of $\bnum 2=\sms{\iota,\iota}\Rightarrow\iota$ is $\estr 2=\lambda x^{\sms{\iota,\iota}}.\pi_\iota x$. Notice that $\lambda x^\iota.\lambda y^\iota.x$ is an inhabitant of $\bnum 2$, however $\lambda x^\iota.\lambda y^\iota.x$ and $\estr 2$ are observationally equivalent.
In general, the only inhabitant of $\bnum n$ are $\estr n= \lambda x^{\sms{\iota,\dots,\iota}}.\pi_\iota x$, and some observationally equivalent terms. That is, the function taking a $n$-tuple and returning non-deterministically one of its elements. If $n$ is bigger than $m$, the non-deterministic choice of $\estr n$ is choosing among more elements than $\estr m$, so we can say the non-deterministic degree is higher.

The interesting feature is that the type is determining the non-determinism degree of the function $\lambda x.\pi_\iota x$. A possible future work is to study this way to distinguish different degrees of non-determinism through types in some other non-deterministic settings such as~\cite{BoudolIC94,BucciarelliEhrhardManzonettoAPAL12,
  deLiguoroPipernoIC95,
  %DezaniciancagliniDeliguoroPipernoSIAM98,
  PaganiRonchidellaroccaFI10,
DiazcaroManzonettoPaganiLFCS13}.

\paragraph{Polymorphism}
An ongoing work is to define a polymorphic version of $\lambda^+$, which will include two more isomorphisms:
\(
  \forall X.(R\wedge S)\equiv\forall X.R\wedge\forall X.S
\),
which is analogous to (dist), and
\(
  \forall X.\forall Y.R\equiv\forall Y.\forall X.R
\),
which is analogous to the combination of the isomorphisms (curry) and (comm) in arrows.

Besides the usefulness of polymorphism for everyday programming, polymorphism can also contribute to the studying of non-determinism mentioned in the previous paragraph: An abstraction $\lambda x.\pi_\iota x$, with $x$ of a polymorphic type could be the generic abstraction to control degrees of non-determinism.

\acks

We would like to thank Ali Assaf, Eduardo Bonelli, Gilles Dowek and Sylvain Salvati for enlightening discussions.

% We recommend abbrvnat bibliography style.

\bibliographystyle{abbrvnat}
\bibliography{biblio}

\begin{thebibliography}{20}
\providecommand{\natexlab}[1]{#1}
\providecommand{\url}[1]{\texttt{#1}}
\expandafter\ifx\csname urlstyle\endcsname\relax
  \providecommand{\doi}[1]{doi: #1}\else
  \providecommand{\doi}{doi: \begingroup \urlstyle{rm}\Url}\fi

\bibitem[Arrighi and D{\'\i}az-Caro(2012)]{ArrighiDiazcaroLMCS12}
P.~Arrighi and A.~D{\'\i}az-Caro.
\newblock A {S}ystem {F} accounting for scalars.
\newblock \emph{Logical Methods in Computer Science}, 8\penalty0 (1:11), 2012.

\bibitem[Arrighi et~al.(2015)Arrighi, D{\'\i}az-Caro, and
  Valiron]{ArrighiDiazcaroValiron13}
P.~Arrighi, A.~D{\'\i}az-Caro, and B.~Valiron.
\newblock The vectorial lambda-calculus.
\newblock Draft at {\tt arXiv:1308.1138}, 2015.
\newblock Considered for publication at Information \& Computation.

\bibitem[Barbanera et~al.(1995)Barbanera, Dezani-Ciancaglini, and de'
  Liguoro]{BarbaneraDezaniDeLiguoroIC95}
F.~Barbanera, M.~Dezani-Ciancaglini, and U.~de' Liguoro.
\newblock {Intersection and Union Types: Syntax and Semantics}.
\newblock \emph{Information and Computation}, 119:\penalty0 202--230, 1995.

\bibitem[Beki\'c(1984)]{BekicLNCS}
H.~Beki\'c.
\newblock Definable operations in general algebras, and the theory of automata
  and flowcharts.
\newblock In \emph{Programming Languages and Their Definition. H. Beki\'c
  (1936--1982)}, volume 177 of \emph{Lecture Notes in Computer Science}, pages
  30--55. Springer, 1984.

\bibitem[Bernadet and Lengrand(2013)]{BernadetLengrandLMCS13}
A.~Bernadet and S.~J. Lengrand.
\newblock Non-idempotent intersection types and strong normalisation.
\newblock \emph{Logical Methods in Computer Science}, 9\penalty0 (4:3), 2013.

\bibitem[Boudol(1994)]{BoudolIC94}
G.~Boudol.
\newblock Lambda-calculi for (strict) parallel functions.
\newblock \emph{Information \& Computation}, 108\penalty0 (1):\penalty0
  51--127, 1994.

\bibitem[Bruce et~al.(1992)Bruce, {Di Cosmo}, and
  Longo]{BruceDiCosmoLongoMSCS92}
K.~B. Bruce, R.~{Di Cosmo}, and G.~Longo.
\newblock Provable isomorphisms of types.
\newblock \emph{Mathematical Structures in Computer Science}, 2\penalty0
  (2):\penalty0 231--247, 1992.

\bibitem[Bucciarelli et~al.(2012)Bucciarelli, Ehrhard, and
  Manzonetto]{BucciarelliEhrhardManzonettoAPAL12}
A.~Bucciarelli, T.~Ehrhard, and G.~Manzonetto.
\newblock A relational semantics for parallelism and non-determinism in a
  functional setting.
\newblock \emph{Annals of Pure and Applied Logic}, 163\penalty0 (7):\penalty0
  918--934, 2012.

\bibitem[Castagna et~al.(2014)Castagna, Nguyen, Xu, Im, Lenglet, and
  Padovani]{CastagnaNguyXuImLengletPadovaniPOPL14}
G.~Castagna, K.~Nguyen, Z.~Xu, H.~Im, S.~Lenglet, and L.~Padovani.
\newblock Polymorphic functions with set-theoretic types: Part 1: Syntax,
  semantics, and evaluation.
\newblock In \emph{Proceedings of the 41st ACM SIGPLAN-SIGACT Symposium on
  Principles of Programming Languages}, POPL '14, pages 5--17, New York, NY,
  USA, 2014. ACM.

\bibitem[De~Benedetti and Ronchi Della~Rocca(2013)]{DebenadettieRonchiITRS12}
E.~De~Benedetti and S.~Ronchi Della~Rocca.
\newblock Bounding normalization time through intersection types.
\newblock In L.~Paolini, editor, \emph{Proceedings of Sixth Workshop on
  Intersection Types and Related Systems (ITRS 2012)}, EPTCS, pages 48--57.
  Cornell University Library, 2013.

\bibitem[de'Liguoro and Piperno(1995)]{deLiguoroPipernoIC95}
U.~de'Liguoro and A.~Piperno.
\newblock Non deterministic extensions of untyped $\lambda$-calculus.
\newblock \emph{Information \& Computation}, 122\penalty0 (2):\penalty0
  149--177, 1995.

\bibitem[Di~Cosmo and Kesner(1993)]{DicosmoKesnerLNCS93}
R.~Di~Cosmo and D.~Kesner.
\newblock A confluent reduction for the extensional typed $\lambda$-calculus
  with pairs, sums, recursion and terminal object.
\newblock In \emph{ICALP'93}, volume 700 of \emph{Lecture Notes in Computer
  Science}, pages 645--656. Springer, 1993.

\bibitem[D{\'\i}az-Caro and Dowek(2015)]{DiazcaroDowek15}
A.~D{\'\i}az-Caro and G.~Dowek.
\newblock Simply typed lambda-calculus modulo type isomorphisms.
\newblock Draft at {\tt hal-01109104}, 2015.

\bibitem[D\'iaz-Caro and Dowek(2016)]{DiazcaroDowek16}
A.~D\'iaz-Caro and G.~Dowek.
\newblock Quantum superpositions and projective measurement in the lambda
  calculus.
\newblock {\tt arXiv:1601.04294}, 2016.

\bibitem[D{\'\i}az-Caro et~al.(2013)D{\'\i}az-Caro, Manzonetto, and
  Pagani]{DiazcaroManzonettoPaganiLFCS13}
A.~D{\'\i}az-Caro, G.~Manzonetto, and M.~Pagani.
\newblock Call-by-value non-determinism in a linear logic type discipline.
\newblock In \emph{LFCS'13}, volume 7734 of \emph{Lecture Notes in Computer
  Science}, pages 164--178, 2013.

\bibitem[Dunfield(2014)]{DunfieldJFP14}
J.~Dunfield.
\newblock Elaborating intersection and union types.
\newblock \emph{J. Functional Programming}, 24\penalty0 (2--3):\penalty0
  133--165, 2014.

\bibitem[Garrigue and A{\"i}t-Kaci(1994)]{GarrigueAitkaciPOPL94}
J.~Garrigue and H.~A{\"i}t-Kaci.
\newblock The typed polymorphic label-selective $\lambda$-calculus.
\newblock In \emph{Proceedings of POPL 1994}, ACM SIGPLAN, pages 35--47, 1994.

\bibitem[Kesner and Ventura(2014)]{KesnerVenturaLNCS14}
D.~Kesner and D.~Ventura.
\newblock Quantitative types for the linear substitution calculus.
\newblock In J.~Diaz, I.~Lanese, and D.~Sangiorgi, editors, \emph{Theoretical
  Computer Science}, volume 8705 of \emph{Lecture Notes in Computer Science},
  pages 296--310. Springer Berlin Heidelberg, 2014.

\bibitem[Pagani and {Ronchi Della Rocca}(2010)]{PaganiRonchidellaroccaFI10}
M.~Pagani and S.~{Ronchi Della Rocca}.
\newblock Linearity, non-determinism and solvability.
\newblock \emph{Fundamenta Informaticae}, 103\penalty0 (1--4):\penalty0
  173--202, 2010.

\bibitem[{P}imentel et~al.(2012){P}imentel, {Ronchi Della Rocca}, and
  {R}oversi]{PimentelRonchiRoversiFI12}
E.~{P}imentel, S.~{Ronchi Della Rocca}, and L.~{R}oversi.
\newblock {I}ntersection {T}ypes from a {P}roof-theoretic {P}erspective.
\newblock \emph{{F}undamenta {I}nformaticae}, 121\penalty0 (1-4):\penalty0
  253---274, 2012.

\end{thebibliography}

% The bibliography should be embedded for final submission.

%\begin{thebibliography}{}
%\softraggedright
%
%\bibitem[Smith et~al.(2009)Smith, Jones]{smith02}
%P. Q. Smith, and X. Y. Jones. ...reference text...
%
%\end{thebibliography}

\clearpage
\appendix

\section{Detailed Proofs}
\xrecap{Lemma}{lem:canShape}
  The canonical form of a type is produced by the following grammar:
  \[
    C := \ms{C_i\Rightarrow\tau}in
  \]
  with the following conventions:
  \[
    \ms{C_i}i0\Rightarrow\tau = \tau
    \qquad
    \qquad
    \ms{C_i}i1=C_1
  \]
\begin{proof}
  We proceed by induction on the structure of types.
  \begin{itemize}
    \item $\can\tau = \tau = \ms{\ms{C}i0\Rightarrow\tau}i1$.
    \item $\can{R\Rightarrow S}=\ms{\can R\uplus \sms{C_i}\Rightarrow\tau}in$, because by the induction hypothesis, $\can S=\ms{C_i\Rightarrow\tau}in$.
    \item $\can{\ms{R_i}in}=\biguplus_{i=1}^n\can{R_i}$. By the induction hypothesis, $\can{R_i}=\ms{C_{ij}\Rightarrow\tau}j{m_i}$, which concludes the case.
      \qedhere
  \end{itemize}
\end{proof}

\xrecap{Lemma}{lem:uniqueNF}
  If $R\equiv S$, then $\can R=\can S$.
\begin{proof}
  We proceed by structural induction on the relation $\equiv$.
    Associativity and commutativity are trivialized by the use of multisets.
  \begin{description}
    \item[$\sms{R,S}\Rightarrow T\equiv R\Rightarrow S\Rightarrow T$:]~\\

      Let $\can T=\ms{C_i\Rightarrow\tau}in$, then
      $\can{S\Rightarrow T}=\ms{\can S\uplus \sms{C_i}\Rightarrow\tau}in$.
      Therefore,
      \begin{align*}
	\can{\sms{R,S}\Rightarrow T} &=\ms{\can{\sms{R,S}}\uplus\sms{C_i}\Rightarrow\tau}in\\
	&=\ms{(\can R\uplus\can S)\uplus\sms{C_i}\Rightarrow\tau}in\\
	&=\ms{\can R\uplus(\can S\uplus\sms{C_i})\Rightarrow\tau}in\\
	&=\can{R\Rightarrow S\Rightarrow T}
      \end{align*}

    \item[$R\Rightarrow\sms{S,T}\equiv\sms{R\Rightarrow S,R\Rightarrow T}$:]~\\

      Let $\can S=\ms{C_i\Rightarrow\tau}in$ and
      $\can T=\ms{D_j\Rightarrow\tau}jm$, then
      \begin{align*}
	\can{R\Rightarrow\sms{S,T}} =~&\ms{\can R\uplus\sms{C_i}\Rightarrow\tau}in\\
	&\uplus\ms{\can R\uplus\sms{D_j}\Rightarrow\tau}jm\\
	=~&\can{R\Rightarrow S}\uplus\can{R\Rightarrow T}\\
	=~&\can{\sms{R\Rightarrow S,R\Rightarrow T}}
      \end{align*}

    \item[$C{[}R{]}\equiv C{[}S{]}$, with $R\equiv S$:]~\\
      
      By the induction hypothesis, $\can R=\can S$. Then,
      \begin{itemize}
	\item $R\Rightarrow T\equiv S\Rightarrow T$. Let $\can T=\ms{C_i\Rightarrow\tau}in$. Then,
	  \begin{align*}
	    \can{R\Rightarrow T} &=\ms{\can R\uplus\sms{C_i}\Rightarrow\tau}in\\
	    &=\ms{\can S\uplus\sms{C_i}\Rightarrow\tau}in\\
	    &=\can{S\Rightarrow T}
	  \end{align*}

	\item $R\Rightarrow S\equiv R\Rightarrow T$. Let $\can S=\can T=\ms{C_i\Rightarrow\tau}in$. Then,
	  \begin{align*}
	    \can{R\Rightarrow S} & =\ms{\can R\uplus\sms{C_i}\Rightarrow\tau}in\\
	    &=\can{R\Rightarrow T}
	  \end{align*}
	\item $\sms{R,S}\equiv\sms{T,S}$.
	  \begin{align*}
	    \can{\sms{R,S}}&=\can R\uplus\can S\\
	    &=\can T\uplus\can S=\can{\sms{T,S}}
	  \end{align*}
	\item $\sms{R,S}\equiv\sms{R,T}$
	  \begin{align*}
	    \can{\sms{R,S}}&=\can R\uplus\can S\\
	    &=\can R\uplus\can T=\can{\sms{R,T}}
	  \end{align*}
	  \qedhere
      \end{itemize}
  \end{description}
\end{proof}

\xrecap{Lemma}{lem:eqCan}
  For any $R$, $R\equiv\can R$.
\begin{proof}
  We proceed by induction on the structure of types.
  \begin{description}
    \item[$\tau$:] Notice that $\can{\tau}=\tau$.
    \item[$R\Rightarrow S$:] By Lemma~\ref{lem:canShape}, $\can S=\ms{C_i\Rightarrow\tau}in$. 
      Then we have $\can{R\Rightarrow S}=\ms{\can R\uplus\sms{C_i}\Rightarrow\tau}in$.
      By the induction hypothesis, $S\equiv\ms{C_i\Rightarrow\tau}in$.
      Hence $R\Rightarrow S\equiv R\Rightarrow\ms{C_i\Rightarrow\tau}in
      \equiv\ms{R\Rightarrow C_i\Rightarrow\tau}in
      \equiv\ms{\can R\uplus\sms{C_i}\Rightarrow\tau}in$.
    \item[$\ms{R_i}in$:] By the induction hypothesis $\can{R_i}\equiv R_i$, hence      
      $\can{\ms{R_i}in}=\biguplus_{i=1}^n\can{R_i}\equiv\biguplus_{i=1}^n R_i=\ms{R_i}in$.
      \qedhere
  \end{description}
\end{proof}

\xrecap{Lemma}{lem:canuncan}
  $\can{\uncan{\ve r}}=\ve r$ and $\uncan{\can{\ve r}}\eq\ve r$.
\begin{proof}
  First we show two properties:\\
  Property 1: $\can{\uncan C}=C$\\
  Property 2: $\uncan{\can R}\equiv R$.\\
  The first propertystatement follows from the fact that $\uncan C$ only changes mutisets by conjunctions, and the canonical of a canonized type is the same type. The second property follows from Lemma~\ref{lem:eqCan}.
  \begin{itemize}
    \item We proceed by induction on $\ve r$ in the implementation of $\lambda^+$.
      \begin{itemize}
	\item $\can{\uncan{x^C}}=\can{x^{\uncan C}}=x^{\can{\uncan C}}=x^C$.
	\item 
	  $\begin{aligned}[t]
	  \can{\uncan{\lambda x^C.\ve r}}&=\can{\lambda x^{\uncan C}.\uncan{\ve r}}\\
	  &=\lambda x^{\can{\uncan C}}.\can{\uncan{\ve r}}
	  \end{aligned}$

	  By the induction hypothesis, $\can{\uncan{\ve r}}=\ve r$, and by the Property 1, $\can{\uncan C}=C$.
	\item $\can{\uncan{\ve r\ve s}}=\can{\uncan{\ve r}}\can{\uncan{\ve r}}$. We conclude by the induction hypothesis.
	\item $\can{\uncan{\ms{\ve r_i}in}}
	  =
	  \can{\sum_{i=1}^n\uncan{\ve r_i}}
	  =
	  \ms{\ve r_i}in
	  $
	\item $
	  \begin{aligned}[t]
	  \can{\uncan{\pi_C(\ve r)}}
	  &=
	  \can{\pi_{\uncan C}(\uncan{\ve r})}\\
	  &=
	  \pi_{\can{\uncan C}}(\can{\uncan{\ve r}})
	  \end{aligned}
	  $

	  By the induction hypothesis, ${\can{\uncan{\ve r}}}=\ve r$, and by the Property 1, $\can{\uncan{C}}=C$.
      \end{itemize}
    \item We proceed by induction on $\ve r$ in $\lambda^+$.
      \begin{itemize}
	\item $\uncan{\can{x^R}}=x^{\uncan{\can R}}\eq x^R$.
	\item $\uncan{\can{\lambda x^R.\ve r}}
	  =\lambda x^{\uncan{\can R}}.\uncan{\can R}$. By the Property 2, $\uncan{\can R}\equiv R$ and by the induction hypothesis, ${\uncan{\can{\ve r}}}\eq\ve r$. Hence, we have that $\lambda x^{\uncan{\can R}}.\uncan{\can R}\eq\lambda x^R.\ve r$.
	\item $\uncan{\can{\ve r\ve s}}=\uncan{\can{\ve r}}\uncan{\can{\ve r}}$. We conclude by the induction hypothesis.
	\item $
	  \begin{aligned}[t]
	  \uncan{\can{\ve r+\ve s}}
	  &=
	  \uncan{\sms{\can r,\can s}}\\
	  &=
	  \uncan{\can{\ve r}}+\uncan{\can{\ve s}}
	  \end{aligned}$

	  We conclude by the induction hypothesis.
	\item $\uncan{\can{\pi_R(\ve r)}}
	  =
	  \pi_{\uncan{\can R}}(\uncan{\can r})$.
	  By the Property 2, $\uncan{\can R}\equiv R$ and by the induction hypothesis, ${\uncan{\can{\ve r}}}\eq\ve r$. Hence, we have that $\pi_{\uncan{\can R}}(\uncan{\can r})\eq\pi_R(\ve r)$.
	  \qedhere
      \end{itemize}
  \end{itemize}
\end{proof}

\xrecap{Theorem}{thm:TypeEq}\conlista
  \begin{enumerate}
    \item 
      If $\Gamma\vdash\ve r:R$ is derivable in $\lambda^+$, then
      \[
	\can{\Gamma}\vdash\canv r:\can{R}
      \] is derivable in the modified system from Figure~\ref{fig:cantypes}.
    \item If $\Gamma\vdash\ve r:C$ is derivable in the modified system from Figure~\ref{fig:cantypes}, then \[
	\uncan\Gamma\vdash\uncan{\ve r}:\uncan C
      \] is derivable in $\lambda^+$.
  \end{enumerate}
\begin{proof}~
  \begin{enumerate}
    \item We proceed by induction on the derivation tree of $\Gamma\vdash\ve r:R$
      \begin{itemize}
	\item Let $\Gamma,x^R\vdash x^R:R$ as a consequence of rule $(\textsl{ax})$. Notice that 
	  \[
	    \infer[^{(\textsl{ax})}]{\can\Gamma,x^{\can R}\vdash x^{\can R}:\can R}{}
	  \]

	\item Let $\Gamma\vdash\ve r:S$ as a consequence of $\Gamma\vdash\ve r:R$ and rule $(\equiv)$. By Lemma~\ref{lem:uniqueNF}, $\can R=\can S$, and by the induction hypothesis, $\can\Gamma\vdash\canv r:\can R=\can S$.

	\item Let $\Gamma\vdash\lambda x^S.\ve r:S\Rightarrow R$ as a consequence of $\Gamma,x^S\vdash\ve r:R$ and rule $({\Rightarrow_i})$.
	  By the induction hypothesis, $\can\Gamma,x^{\can S}\vdash\canv r:\can R$. 
	  Let $\can R=\ms{C_i\Rightarrow\tau}in$.
	  Hence,

	  \scalebox{0.89}{
	    \parbox{1.08\linewidth}{
	      \[
		\infer[^{(\Rightarrow_i)}]{\can\Gamma\vdash\lambda x^{\can S}.\canv r:
		\ms{\can S\uplus\sms{C_i}\Rightarrow\tau}in}
		{\can\Gamma,x^{\can S}\vdash\canv r:\ms{C_i\Rightarrow\tau}in}
	      \]
	    }
	  }

	  Notice that 
	  \begin{align*}
	    \ms{\can S\uplus\sms{C_i}\Rightarrow\tau}in &= \can{S\Rightarrow R}\\
	    \lambda x^{\can S}.\canv r &=\can{\lambda x^S.\ve r}
	  \end{align*}

	\item Let $\Gamma\vdash\ve r\ve s:R$ as a consequence of $\Gamma\vdash\ve r:S\Rightarrow R$, $\Gamma\vdash\ve s:S$ and rule ${(\Rightarrow_e)}$.
	  By the induction hypothesis, $\can\Gamma\vdash\canv r:\can{S\Rightarrow R}$
	  and $\can\Gamma\vdash\canv s:\can S$.

	  By Lemma~\ref{lem:canShape}, 
	  \[
	    \can R=\ms{\ms{C_j}jn\Rightarrow\tau}km
	  \]
	  Hence, by definition~\ref{def:canonical},
	  \[
	    \can{S\Rightarrow R}=\ms{\can S\uplus\ms{C_j}jn\Rightarrow\tau}km
	  \]
	  Then,

	  \scalebox{0.94}{
	    \parbox{1.04\linewidth}{
	      \[
		\infer[^{(\Rightarrow_e)}]{\can\Gamma\vdash\canv{\ve rs}:\ms{\ms{C_j}jn\Rightarrow\tau}km}
		{
		  \begin{array}{c}
		    \can\Gamma\vdash\canv r:\ms{\can S\uplus\ms{C_j}jn\Rightarrow\tau}km\hspace{1.2cm}\\
		    \hfill \can\Gamma\vdash\canv s:\can S
		  \end{array}
		}
	      \]
	    }
	  }

	  Notice that $\ms{\ms{C_j}jn\Rightarrow\tau}km=\can R$.

	\item Let $\Gamma\vdash\ve r_1+\ve r_2:R_1\wedge R_2$ as a consequence of $\Gamma\vdash\ve r_i:R_i$, with $i=1,2$, and rule $(\wedge_i)$.
	  By the induction hypothesis, $\can\Gamma\vdash\canv r_i:\can R_i$, for $i=1,2$.
	  Hence,

	  \scalebox{0.94}{
	    \parbox{1.04\linewidth}{
	  \[
	    \infer[^{(\wedge_i)}]{\can\Gamma\vdash\sms{\can{\ve r_1},\can{\ve r_2}}:\can{R_1}\uplus\can{R_2}}
	    {
	      (i=1,2)\quad 
	      \can\Gamma\vdash\can{\ve r_i}:\can{R_i}
	    }
	  \]
	}
      }

      Notice that $\sms{\can{\ve r_1}+\can{\ve r_2}}=\can{\ve r_1+\ve r_2}$ and $\can{R_1}\uplus\can{R_2}=\can{R_1\wedge R_2}$.

	\item Let $\Gamma\vdash\pi_R(\ve r):R$ as a consequence of $\Gamma\vdash\ve r:R\wedge S$ and rule $(\wedge_{e_n})$.
	  By the induction hypothesis,
	  $\can\Gamma\vdash\canv r:\can{R\wedge S}$. Notice that $\can{R\wedge S}=\sms{\can R,\can S}$. Then
	  \[
	    \infer[^{(\wedge_e)}]{\can\Gamma\vdash\pi_{\can R}(\canv r):\can R}
	    {\can\Gamma\vdash\canv r:\sms{\can R,\can S}}
	  \]
	  Notice that $\can{\pi_R(\ve r)}=\pi_{\can R}(\canv r)$.

	\item Let $\Gamma\vdash\pi_R(\ve r):R$ as a consequence of $\Gamma\vdash\ve r:R$ and rule $(\wedge_{e_1})$.
	  By the induction hypothesis,
	  $\can\Gamma\vdash\canv r:\can R$. Then
	  \[
	    \infer[^{(\wedge_e)}]{\can\Gamma\vdash\pi_{\can R}(\canv r):\can R}
	    {\can\Gamma\vdash\canv r:\can R}
	  \]
	  Notice that $\can{\pi_R(\ve r)}=\pi_{\can R}(\canv r)$.
      \end{itemize}
    \item We proceed by induction on the derivation three of $\Gamma\vdash\ve t:C$.
      \begin{itemize}
	\item Let $\Gamma,x^{\gms C}\vdash x^{\gms C}:\gms C$ as a consequence of rule $(\textsl{ax})$. Then

	  \scalebox{0.96}{
	    \parbox{1.01\linewidth}{
	      \[
		\infer[^{(\textsl{ax})}]{\uncan\Gamma,x^{\uncan{\gms C}}\vdash x^{\uncan{\gms C}}:\uncan{\gms C}}
		{}
	      \]
	    }
	  }

	\item Let $\Gamma\vdash\lambda x^{\gms C}.\ve r:\ms{\gms C\uplus\sms{D_i}\Rightarrow\tau}in$ as a consequence of $\Gamma,x^{\gms C}\vdash\ve r:\ms{D_i\Rightarrow\tau}in$ and rule $(\Rightarrow_i)$. Then by the induction hypothesis, we have $\uncan\Gamma,x^{\uncan{\gms C}}\vdash\uncan{\ve r}:\bigwedge_{i=1}^n{\uncan{D_i}\Rightarrow\tau}$.
	  Notice that 

	  \scalebox{0.9}{
	    \parbox{1.09\linewidth}{
	      \[
		\uncan{\gms C}\Rightarrow\ws{\uncan{D_i}\Rightarrow\tau}in
		\equiv
		\ws{\uncan{\gms C\uplus\sms{D_i}}\Rightarrow\tau}in
	      \]
	    }
	  }

	  Hence,

	  \scalebox{0.75}{
	    \parbox{1.3\linewidth}{
	      \[
		\infer[^{(\equiv)}]{\uncan\Gamma\vdash\lambda x^{\uncan{\gms C}}.\uncan{\ve r}:
		\ws{\uncan{\gms C\uplus\sms{D_i}}\Rightarrow\tau}in}
		{
		  \infer[^{(\Rightarrow_i)}]{\uncan\Gamma\vdash\lambda x^{\uncan{\gms C}}.\uncan{\ve r}:\uncan{\gms C}\Rightarrow\ws{\uncan{D_i}\Rightarrow\tau}in}
		  {
		    \uncan\Gamma,x^{\uncan{\gms C}}\vdash\uncan{\ve r}:\ws{\uncan{D_i}\Rightarrow\tau}in
		  }
		}
	      \]
	    }
	  }

	  Notice that $\lambda x^{\uncan{\gms C}}.\uncan{\ve r}=\uncan{\lambda x^{\gms C}.\ve r}$, and\\ $\ws{\uncan{\gms C\uplus\sms{D_i}}\Rightarrow\tau}in=\uncan{\ms{\gms C\uplus\sms{D_i}\Rightarrow\tau}in}$.

	\item Let $\Gamma\vdash\ve r\ve s:\ms{\gms C_k\setminus\gms D\Rightarrow\tau}km$ as a consequence of
	  $\Gamma\vdash\ve r:\ms{\gms C_k\Rightarrow\tau}km$,
	  $\Gamma\vdash\ve s:\gms D$,
	  $\gms D\subseteq\bigcap_{k=1}^m\gms C_k$
	  and rule ${(\Rightarrow_e)}$.
	  Then by the induction hypothesis, 
	  $\uncan\Gamma\vdash\uncan{\ve r}:\ws{\uncan{\gms C_k}\Rightarrow\tau}km$,
	  and\\
	  $\uncan\Gamma\vdash\uncan{\ve s}:\uncan{\gms D}$.
	  Notice that 
	  \begin{align*}
	    &\ws{\uncan{\gms C_k}\Rightarrow\tau}km \\
	    &\equiv
	    \ws{\uncan{\gms D}\Rightarrow\uncan{\gms C_k\setminus\gms D}\Rightarrow\tau}km \\ 
	    &\equiv
	    \uncan{\gms D}\Rightarrow\ws{\uncan{\gms C_k\setminus\gms D}\Rightarrow\tau}km 
	  \end{align*}
	  Hence,

	  \scalebox{0.9}{
	    \parbox{1.08\linewidth}{
	  \[
	    \uncan\Gamma\vdash\uncan{\ve r}:\uncan{\gms D}\Rightarrow\ws{\uncan{\gms C_k\setminus\gms D}\Rightarrow\tau}km\
	  \]
	}
      }

	  and since
	  \[
	    \uncan\Gamma\vdash\uncan{\ve s}:\uncan{\gms D}
	  \]
	  by rule $(\Rightarrow_e)$, we have
	  \[
	    \uncan\Gamma\vdash\uncan{\ve r}\uncan{\ve s}:\ws{\uncan{\gms C_k\setminus\gms D}\Rightarrow\tau}km
	  \]

	  Notice that $\uncan{\ve r}\uncan{\ve s}=\uncan{\ve r\ve s}$ and
	  \begin{align*}
	    &\ws{\uncan{\gms C_k\setminus\gms D}\Rightarrow\tau}km\\
	    &= \uncan{\ms{\gms C_k\setminus\gms D\Rightarrow\tau}km}
	  \end{align*}

	\item Let $\Gamma\vdash\ms{\ve r_i}in:\ms{C_i}in$ as a consequence of $\Gamma\vdash\ve r_i:\sms{C_i}$, for $i=1,\dots,n$, and rule $(\wedge_i)$. Then by the induction hypothesis, $\uncan\Gamma\vdash\uncan{\ve r_i}:\uncan{C_i}$, for $i=1,\dots,n$. Hence,

	  \scalebox{0.7}{
	    \parbox{1.4\linewidth}{
	      \[
		\infer[^{(\wedge_i)}]{\uncan\Gamma\vdash\sum\limits_{i=1}^n\uncan{\ve r_i}:\wsl{\uncan{C_i}}in}
		{
		  \uncan\Gamma\vdash\uncan{\ve r_1}:\uncan{C_1}
		  &
		  \infer[^{(\wedge_i)}]{\uncan\Gamma\vdash\sum\limits_{i=2}^n\uncan{\ve r_i}:\bigwedge\limits_{i=2}^n{\uncan{C_i}}}
		  {\vdots}
		}
	      \]
	    }
	  }

	  Notice that $\sum\limits_{i=1}^n\uncan{\ve r_i}=\uncan{\ms{\ve r_i}in}$.

	\item Let $\Gamma\vdash\pi_{\gms D}\ve r:\gms D$ as a consequence of $\Gamma\vdash\ve r:\gms C$ and rule $(\wedge_e)$, with $\gms D\subseteq\gms C$. Then by the induction hypothesis,
	  $\uncan\Gamma\vdash\uncan{\ve r}:\uncan{\gms C}$. Hence, either by $(\wedge_{e_n})$ or $(\wedge_{e_1})$, depending if $\gms D=\gms C$ or $\gms D\subset\gms C$, we have
	  \[
	    \infer[^{(\wedge_e)}]{\uncan\Gamma\vdash\pi_{\uncan{\gms D}}\uncan{\ve r}:\uncan{\gms D}}{\uncan\Gamma\vdash\uncan{\ve r}:\uncan{\gms C}}
	  \]

	  Notice that $\pi_{\uncan{\gms D}}\uncan{\ve r}=\uncan{\pi_{\gms D}(\ve r)}$.
	  \qedhere
      \end{itemize}
  \end{enumerate}
\end{proof}

\xrecap{Theorem}{thm:rewSoundComp}
Let $\ve r$ be a closed term in $\lambda^+$.
\begin{itemize}
  \item If $\ve r\eq\ve r'$ by any rule other than AC, then 
    \begin{itemize}
      \item 
	if $\vdash\ve r:R\Rightarrow S$, then for all $\vdash\ve s:\can R$, where $\ve s$ is in canonical form, there exists $\ve t$ such that
	\begin{center}
	  \begin{tikzpicture}
	    \node (u) at (0,0) {$\can{\ve r}\ve s$};
	    \node (d) at (2,0) {$\can{\ve r'}\ve s$};
	    \node (t) at (1,-1) {$\ve t$};	    
	    \path (u) edge[to*] (t) (d) edge[to*] (t);
	  \end{tikzpicture}
	\end{center}
      \item Otherwise, there exists $\ve t$ such that 
	\begin{center}
	  \begin{tikzpicture}
	    \node (u) at (0,0) {$\can{\ve r}$};
	    \node (d) at (2,0) {$\can{\ve r'}$};
	    \node (t) at (1,-1) {$\ve t$};	    
	    \path (u) edge[to*] (t) (d) edge[to*] (t);
	  \end{tikzpicture}
	\end{center}
    \end{itemize}
  \item If $\ve r\hookrightarrow\ve r'$, then $\can{\ve r}\to^+\can{\ve r'}$.
\end{itemize}
\begin{proof}
  We proceed by checking rule by rule. Notice that if there exists $\ve t$ such that $\can{\ve r}\to^*\ve t$, then there also exists $\ve t\ve s$ such that $\can{\ve r}\ve s\to^*\ve t\ve s$.
  \begin{description}
    \item[$\lambda x^R.(\ve r+\ve r')\eq\lambda x^R.\ve r+\lambda x^R.\ve r'$:]~\\
      Let $\can R=\gms C$ and $\vdash\ve s:\gms C$. Then,

      \begin{center}
	\begin{tikzpicture}
	  \node (u) at (-1.5,2.5) {$\can{\lambda x^R.(\ve r+\ve r')}\ve s$};
	  \node (d) at (-1.5,1.5) {$(\lambda x^{\gms C}.\sms{\canv r,\can{\ve r'}})\ve s$};
	  \node (t) at (0,0) {$\sms{\canv r[\ve s/x],\can{\ve r'}[\ve s/x]}$};
	  \node (c) at (1.5,3) {$\can{\lambda x^R.\ve r+\lambda x^R.\ve r'}\ve s$};
	  \node (ci) at (1.5,2) {$\sms{\lambda x^{\gms C}.\canv r,\lambda x^{\gms C}.\can{\ve r'}}\ve s$};
	  \node (s) at (1.5,1) {$\sms{(\lambda x^{\gms C}.\canv r)\ve s,(\lambda x^{\gms C}.\can{\ve r'})\ve s}$};
	  \path (u) edge[double] (d)
	  (d) edge[bend right,->] (t)
	  (c) edge[double] (ci)
	  (ci) edge[->] (s)
	  (s) edge[to*] (t);
	\end{tikzpicture}
      \end{center}

    \item[$(\ve r+\ve s)\ve t\eq\ve r\ve t+\ve s\ve t$:]~
      \begin{align*}
      \can{(\ve r+\ve s)\ve t} &=\sms{\can{\ve r},\can{\ve s}}\can{\ve t}\\
      &\to\sms{\can{\ve r}\can{\ve t},\can{\ve s}\can{\ve t}}\\
      &=\can{\ve r\ve t+\ve s\ve t}
      \end{align*}

    \item[$\pi_{R\Rightarrow S}(\lambda x^R.\ve r)\eq\lambda x^R.\pi_S(\ve r)$:]~\\
      Let $\can S=\ms{C_i\Rightarrow\tau}in$, hence $\can{R\Rightarrow S}=\ms{\can R\uplus\sms{C_i}\Rightarrow\tau}in$.
      So,
      \begin{align*}
	&\can{\pi_{R\Rightarrow S}(\lambda x^R.\ve r)}\\
	&=\pi_{\ms{\can R\uplus\sms{C_i}\Rightarrow\tau}in}(\lambda x^{\can R}.\can{\ve r})\\
	&\to\lambda x^{\can R}.\pi_{\ms{C_i\Rightarrow\tau}in}(\can{\ve r})\\
	&=\can{\lambda x^R.\pi_S(\ve r)}
      \end{align*}

    \item[$\pi_{S\Rightarrow R}(\ve r)\ve s\eq\pi_R(\ve r\ve s)$] with $\Gamma\vdash\ve r:S\Rightarrow(R\wedge T)$ and $\Gamma\vdash\ve s:S$\textbf{:}~\\
      Let $\can S=\gms D$ and $\can R=\ms{C_i\Rightarrow\tau}in$. Then we have $\can{S\Rightarrow R}=\ms{\gms D\uplus\gms{C_i}\Rightarrow\tau}in$. Notice that,
      $\pi_{\ms{C_i\Rightarrow\tau}in}(\ve r\ve s)\to\pi_{\ms{\gms D\uplus\gms{C_i}\Rightarrow\tau}in}(\ve r)\ve s$.

    \item[$\ve r\ve s\ve t\eq\ve r(\ve s+\ve t)$:]~
      \begin{align*}
	\can{\ve r(\ve s+\ve t)}
        &=\canv r\sms{\canv s,\canv t}\\
	&\to\canv r\canv s\canv t\\
	&=\canv{rst}
      \end{align*}

    \item[$\ve r\eq\ve r\repl{R/S}$] with $R\equiv S$\textbf{:}~\\
      By Lemma~\ref{lem:uniqueNF}, $\can R=\can S$, so
      \begin{align*}
      \canv r &=\can{\ve r\repl{\can S/S}}\\
      &=\can{\ve r\repl{\can R/S}}\\
      &=\can{\ve r\repl{R/S}}
      \end{align*}

    \item[$\pi_{R\wedge S}(\ve r+\ve s)\eq\pi_R(\ve r)+\pi_S(\ve s)$] with $\Gamma\vdash\ve r:R\wedge R'$ or $\Gamma\vdash\ve r:R$, and $\Gamma\vdash\ve s:S\wedge S'$ or $\Gamma\vdash\ve s:S$\textbf{:}~\\
      Let $\can R=\gms C$, $\can{R'}=\gms{C'}$, $\can S=\gms D$ and $\can{S'}=\gms{D'}$. Then,
      \begin{align*}
	\can{\pi_{R\wedge S}(\ve r+\ve s)}
	&=\pi_{\gms C\uplus\gms D}\sms{\canv r,\canv s}\\
	&\to\sms{\pi_{\gms C}(\canv r),\pi_{\gms D}(\canv s)}\\
	&=\can{\pi_R(\ve r)+\pi_S(\ve s)}
      \end{align*}

    \item[$(\lambda x^R.\ve r)\ve s\hookrightarrow\ve r\repl{\ve s/x}$] with $\Gamma\vdash\ve s:R$\textbf{:}~\\
      Let $\can R=\gms C$, then
      \begin{align*}
	\can{(\lambda x^R.\ve r)\ve s}
	&=(\lambda x^{\gms C}.\canv r)\canv s\\
	&\to\canv r[\canv s/x]\\
	&=\can{\ve r[\ve s/x]}
      \end{align*}

    \item[$\pi_R(\ve r+\ve s)\hookrightarrow\ve r$] with $\Gamma\vdash\ve r:R$\textbf{:}~\\
      Let $\can R=\gms C$. Then
      \begin{align*}
	\can{\pi_R(\ve r+\ve s)}
	&=\pi_{\gms C}\sms{\canv r,\canv s}\\
	&\to\pi_{\gms C}(\canv r)\\
	&\to\canv r
      \end{align*}

    \item[$\pi_R(\ve r)\hookrightarrow\ve r$] with $\Gamma\vdash\ve r:R$\textbf{:}~\\
      Let $\can R=\gms C$, then
      \[
	\can{\pi_R(\ve r)}=\pi_{\gms C}(\canv r)\to\canv r
      \]

    \item[$\ve r\hookrightarrow\pi_{R}(\ve r)+\pi_S(\ve r)$]
    with $\Gamma\vdash\ve r:R\wedge S$, $\ve r\not\eq^*\ve s+\ve t$ with
  $\Gamma\vdash\ve s:R$ and $\Gamma\vdash\ve t:S$\textbf{:}~\\
      Let $\can R=\gms C$ and $\can S=\gms D$, then
      \begin{align*}
	\can{\ve r}
	&\to\sms{\pi_{\gms C}(\can{\ve r}),\pi_{\gms D}(\can{\ve r})}\\
	&=\can{\pi_R(\ve r)+\pi_S(\ve r)}
      \qedhere
      \end{align*}
  \end{description}
\end{proof}

\newpage
\section{Trace of $\s{div}$}\label{ap:trace}
\begin{supertabular}{*{50}{p{2mm}@{}}}
  \m{50}{\underline{\s{div}}}\\
  \expl={Definition of \s{div}}\\
  \all{\underline{\fstO{Nat\times Nat\Rightarrow Nat}(\divMod 34)}}\\[1em]
  \expl={Definition of \fstO{}}\\
  \all{(\pi_{Nat\times Nat\Rightarrow\cnat 1}(\underline{\divMod 34}))\estr 1}\\[1em]
  \expl={Definition of \divMod{}{}}\\
  \all{(\underline{\pi_{Nat\times Nat\Rightarrow\cnat 1}}}\\
  &\m{49}{(\underline{\lambda x^{Nat\times Nat}.}}\\
  &&\m{48}{\divModRec 34\canon{\fst(x)}3\canon{\snd(x)}4 0}\\
  &\m{49}{)}\\
  \all{)\estr 1}\\[1em]
  \expl\to{Rule comm$_{ei}$}\\
  \all{(\lambda x^{Nat\times Nat}.}\\
  &\m{49}{\pi_{\cnat 1}}\\
  &&\m{48}{(\underline{\divModRec 34\canon{\fst(x)}3\canon{\snd(x)}4 0})}\\
  \all{)\estr 1}\\[1em]
  \expl={Definition of \divModRec{}{}}\\
  \all{(\lambda x^{Nat\times Nat}.}\\
  &\m{49}{\underline{\pi_{\cnat 1}}}\\
  &&\m{48}{\underline{(}}\\
  &&&\m{47}{\underline{(\mu x^{\sms{\cnat 3,\cnat 4,Nat}\Rightarrow Nat\times Nat}.}}\\
  &&&&\m{46}{\underline{\lambda n^{\cnat 3}.\lambda m^{\cnat 4}.\lambda k^{Nat}.}}\\
  &&&&&\m{45}{\underline{\ifZ~(n\estr 3)}}\\
  &&&&&&\m{44}{\underline{(0,k)}}\\
  &&&&&&\m{44}{\underline{(\ifEq~(m\estr 4)~(\suc k)}}\\
  &&&&&&&\m{43}{\underline{\succFst(x\canon{\pred{(n\estr 3)}}3m0)}}\\
  &&&&&&&\m{43}{\underline{(x\canon{\pred{(n\estr 3)}}3m(\suc k)))}}\\
  &&&\m{47}{\underline{)}}\\
  &&&\m{47}{\underline{\canon{\fst(x)}3}}\\
  &&&\m{47}{\underline{\canon{\snd(x)}4}}\\
  &&&\m{47}{\underline{0}}\\
  &&\m{48}{\underline{)}}\\
  \all{)\estr 1}\\[1em]
  \expl{\to^3}{Rule comm$_{ee}$ ($\times 3$)}\\
  \all{(\lambda x^{Nat\times Nat}.}\\
  &\m{49}{\underline{\pi_{\sms{\cnat 3,\cnat 4,Nat}\Rightarrow\cnat 1}}}\\
  &&\m{48}{(\underline{\mu x^{\sms{\cnat 3,\cnat 4,Nat}\Rightarrow Nat\times Nat}.}}\\
  &&&\m{47}{\lambda n^{\cnat 3}.\lambda m^{\cnat 4}.\lambda k^{Nat}.}\\
  &&&&\m{46}{\ifZ~(n\estr 3)}\\
  &&&&&\m{45}{(0,k)}\\
  &&&&&\m{45}{(\ifEq~(m\estr 4)~(\suc k)}\\
  &&&&&&\m{44}{\succFst(x\canon{\pred{(n\estr 3)}}3m0)}\\
  &&&&&&\m{44}{(x\canon{\pred{(n\estr 3)}}3m(\suc k)))}\\
  &&\m{48}{)}\\
  &\m{49}{\canon{\fst(x)}3}\\
  &\m{49}{\canon{\snd(x)}4}\\
  &\m{49}{0}\\
  \all{)\estr 1}\\[1em]
  \expl\to{Rule comm$_\mu$}\\
  \all{(\lambda x^{Nat\times Nat}.}\\
  &\m{49}{\mu x_1^{\sms{\cnat 3,\cnat 4,Nat}\Rightarrow\cnat 1}.}\\
  &&\m{48}{\underline{\pi_{\sms{\cnat 3,\cnat 4,Nat}\Rightarrow\cnat 1}}}\\
  &&&\m{47}{\underline{\lambda n^{\cnat 3}.\lambda m^{\cnat 4}.\lambda k^{Nat}.}}\\
  &&&&\m{46}{\ifZ~(n\estr 3)}\\
  &&&&&\m{45}{(0,k)}\\
  &&&&&\m{45}{(\ifEq~(m\estr 4)~(\suc k)}\\
  &&&&&&\m{44}{\succFst(
    (\lbrack x_1,R_2\rbrack) %
    \canon{\pred{(n\estr 3)}}3m0
  )}\\
  &&&&&&\m{44}{(
    (\lbrack x_1,R_2\rbrack) %
    \canon{\pred{(n\estr 3)}}3m(\suc k)
  ))}\\
  &&\m{48}{)}\\
  &\m{49}{\canon{\fst(x)}3}\\
  &\m{49}{\canon{\snd(x)}4}\\
  &\m{49}{0}\\
  \all{)\estr 1}\\[1ex]
  \all{\mbox{where }R_2\mbox{ is}}\\
  &&&&&\m{45}{\mu x_2^{\sms{\cnat 3,\cnat 4,Nat}\Rightarrow\cnat 2}.}\\
  &&&&&&\m{44}{\pi_{\sms{\cnat 3,\cnat 4,Nat}\Rightarrow\cnat 2}}\\
  &&&&&&&\m{43}{\lambda n^{\cnat 3}.\lambda m^{\cnat 4}.\lambda k^{Nat}.}\\
  &&&&&&&&\m{42}{\ifZ~(n\estr 3)}\\
  &&&&&&&&&\m{41}{(0,k)}\\
  &&&&&&&&&\m{41}{(\ifEq~(m\estr 4)~(\suc k)}\\
  &&&&&&&&&&\m{40}{\succFst(
    (\lbrack x_1,x_2\rbrack) %
    \canon{\pred{(n\estr 3)}}3m0
  )}\\
  &&&&&&&&&&\m{40}{(
    (\lbrack x_1,x_2\rbrack) %
    \canon{\pred{(n\estr 3)}}3m(\suc k)
  ))}\\[1em]
  \expl{\to^3}{Rule comm$_{ei}$ ($\times 3$)}\\
  \all{(\lambda x^{Nat\times Nat}.}\\
  &\m{49}{\mu x_1^{\sms{\cnat 3,\cnat 4,Nat}\Rightarrow\cnat 1}.}\\
  &&\m{48}{\lambda n^{\cnat 3}.\lambda m^{\cnat 4}.\lambda k^{Nat}.}\\
  &&&\m{47}{\underline{\pi_{\cnat 1}}}\\
  &&&&\m{46}{(\underline{\ifZ~(n\estr 3)}}\\
  &&&&&\m{45}{(0,k)}\\
  &&&&&\m{45}{(\ifEq~(m\estr 4)~(\suc k)}\\
  &&&&&&\m{44}{\succFst(
    (\lbrack x_1,R_2\rbrack) %
    \canon{\pred{(n\estr 3)}}3m0
  )}\\
  &&&&&&\m{44}{(
    (\lbrack x_1,R_2\rbrack) %
    \canon{\pred{(n\estr 3)}}3m(\suc k)
  ))}\\
  &&&&\m{46}{)}\\
  &&\m{48}{)}\\
  &\m{49}{\canon{\fst(x)}3}\\
  &\m{49}{\canon{\snd(x)}4}\\
  &\m{49}{0}\\
  \all{)\estr 1}\\[1em]
  \expl\to{Rule comm$_{\textsl{ifZ}}$}\\
  \all{(\lambda x^{Nat\times Nat}.}\\
  &\m{49}{\mu x_1^{\sms{\cnat 3,\cnat 4,Nat}\Rightarrow\cnat 1}.}\\
  &&\m{48}{\lambda n^{\cnat 3}.\lambda m^{\cnat 4}.\lambda k^{Nat}.}\\
  &&&\m{47}{\ifZ~(n\estr 3)}\\
  &&&&\m{46}{\underline{\pi_{\cnat 1}(0,k)}}\\
  &&&&\m{46}{\pi_{\cnat 1}}\\
  &&&&&\m{45}{(\ifEq~(m\estr 4)~(\suc k)}\\
  &&&&&&\m{44}{\succFst(
    (\lbrack x_1,R_2\rbrack) %
    \canon{\pred{(n\estr 3)}}3m0
  )}\\
  &&&&&&\m{44}{(
    (\lbrack x_1,R_2\rbrack) %
    \canon{\pred{(n\estr 3)}}3m(\suc k)
  )}\\
  &&&&&\m{45}{)}\\
  &&\m{48}{)}\\
  &\m{49}{\canon{\fst(x)}3}\\
  &\m{49}{\canon{\snd(x)}4}\\
  &\m{49}{0}\\
  \all{)\estr 1}\\[1em]
  \expl{\to^2}{Rules simpl and proj}\\
  \all{(\lambda x^{Nat\times Nat}.}\\
  &\m{49}{\mu x_1^{\sms{\cnat 3,\cnat 4,Nat}\Rightarrow\cnat 1}.}\\
  &&\m{48}{\lambda n^{\cnat 3}.\lambda m^{\cnat 4}.\lambda k^{Nat}.}\\
  &&&\m{47}{\ifZ~(n\estr 3)}\\
  &&&&\m{46}{\canon 01}\\
  &&&&\m{46}{\underline{\pi_{\cnat 1}}}\\
  &&&&&\m{45}{(\underline{\ifEq}~(m\estr 4)~(\suc k)}\\
  &&&&&&\m{44}{{\succFst(
    (\lbrack x_1,R_2\rbrack) %
    \canon{\pred{(n\estr 3)}}3m0
  )}}\\
  &&&&&&\m{44}{(
    (\lbrack x_1,R_2\rbrack) %
    \canon{\pred{(n\estr 3)}}3m(\suc k)
  )}\\
  &&&&&\m{45}{)}\\
  &&\m{48}{)}\\
  &\m{49}{\canon{\fst(x)}3}\\
  &\m{49}{\canon{\snd(x)}4}\\
  &\m{49}{0}\\
  \all{)\estr 1}\\[1em]
  \expl\to{Rule comm$_{\textsl{ifEq}}$}\\
  \all{(\lambda x^{Nat\times Nat}.}\\
  &\m{49}{\mu x_1^{\sms{\cnat 3,\cnat 4,Nat}\Rightarrow\cnat 1}.}\\
  &&\m{48}{\lambda n^{\cnat 3}.\lambda m^{\cnat 4}.\lambda k^{Nat}.}\\
  &&&\m{47}{\ifZ~(n\estr 3)}\\
  &&&&\m{46}{\canon 01}\\
  &&&&\m{46}{({\ifEq}~(m\estr 4)~(\suc k)}\\
  &&&&&\m{45}{\pi_{\cnat 1}}\\
  &&&&&&\m{44}{({\succFst(
    (\lbrack x_1,R_2\rbrack) %
    \canon{\pred{(n\estr 3)}}3m0
  )})}\\
  &&&&&\m{45}{\pi_{\cnat 1}}\\
  &&&&&&\m{44}{\underline{((\lbrack x_1,R_2\rbrack)\canon{\pred{(n\estr 3)}}3m(\suc k))}}\\
  &&&&\m{46}{)}\\
  &&\m{48}{)}\\
  &\m{49}{\canon{\fst(x)}3}\\
  &\m{49}{\canon{\snd(x)}4}\\
  &\m{49}{0}\\
  \all{)\estr 1}\\[1em]
  \expl{\to^2}{Rule dist$_i$ ($\times 2$)}\\
  \all{(\lambda x^{Nat\times Nat}.}\\
  &\m{49}{\mu x_1^{\sms{\cnat 3,\cnat 4,Nat}\Rightarrow\cnat 1}.}\\
  &&\m{48}{\lambda n^{\cnat 3}.\lambda m^{\cnat 4}.\lambda k^{Nat}.}\\
  &&&\m{47}{\ifZ~(n\estr 3)}\\
  &&&&\m{46}{\canon 01}\\
  &&&&\m{46}{({\ifEq}~(m\estr 4)~(\suc k)}\\
  &&&&&\m{45}{\pi_{\cnat 1}}\\
  &&&&&&\m{44}{({\succFst(
    (\lbrack x_1,R_2\rbrack) %
    \canon{\pred{(n\estr 3)}}3m0
  )})}\\
  &&&&&\m{45}{\underline{\pi_{\cnat 1}}}\\
  &&&&&&\m{44}{\underline{(\lbrack x_1\canon{\pred{(n\estr 3)}}3m(\suc k),}}\\
  &&&&&&\m{44}{\underline{R_2\canon{\pred{(n\estr 3)}}3m(\suc k)\rbrack)}}\\
  &&&&\m{45}{)}\\
  &&\m{48}{)}\\
  &\m{49}{\canon{\fst(x)}3}\\
  &\m{49}{\canon{\snd(x)}4}\\
  &\m{49}{0}\\
  \all{)\estr 1}\\[1em]
  \expl{\to^2}{Rules simpl and proj}\\
  \all{(\lambda x^{Nat\times Nat}.}\\
  &\m{49}{\mu x_1^{\sms{\cnat 3,\cnat 4,Nat}\Rightarrow\cnat 1}.}\\
  &&\m{48}{\lambda n^{\cnat 3}.\lambda m^{\cnat 4}.\lambda k^{Nat}.}\\
  &&&\m{47}{\ifZ~(n\estr 3)}\\
  &&&&\m{46}{\canon 01}\\
  &&&&\m{46}{({\ifEq}~(m\estr 4)~(\suc k)}\\
  &&&&&\m{45}{\pi_{\cnat 1}}\\
  &&&&&&\m{44}{({\succFst(\underline{(\lbrack x_1,R_2\rbrack)\canon{\pred{(n\estr 3)}}3m0})})}\\
  &&&&&\m{45}{(x_1\canon{\pred{(n\estr 3)}}3m(\suc k))}\\
  &&&&\m{46}{)}\\
  &&\m{48}{)}\\
  &\m{49}{\canon{\fst(x)}3}\\
  &\m{49}{\canon{\snd(x)}4}\\
  &\m{49}{0}\\
  \all{)\estr 1}\\[1em]
  \expl\to{Rule dist$_i$}\\
  \all{(\lambda x^{Nat\times Nat}.}\\
  &\m{49}{\mu x_1^{\sms{\cnat 3,\cnat 4,Nat}\Rightarrow\cnat 1}.}\\
  &&\m{48}{\lambda n^{\cnat 3}.\lambda m^{\cnat 4}.\lambda k^{Nat}.}\\
  &&&\m{47}{\ifZ~(n\estr 3)}\\
  &&&&\m{46}{\canon 01}\\
  &&&&\m{46}{({\ifEq}~(m\estr 4)~(\suc k)}\\
  &&&&&\m{45}{\pi_{\cnat 1}}\\
  &&&&&&\m{44}{\underline{\succFst}}\\
  &&&&&&&\m{43}{\lbrack x_1\canon{\pred{(n\estr 3)}}3m0,R_2\canon{\pred{(n\estr 3)}}3m0\rbrack}\\
  &&&&&\m{45}{(x_1\canon{\pred{(n\estr 3)}}3m(\suc k))}\\
  &&&&\m{46}{)}\\
  &&\m{48}{)}\\
  &\m{49}{\canon{\fst(x)}3}\\
  &\m{49}{\canon{\snd(x)}4}\\
  &\m{49}{0}\\
  \all{)\estr 1}\\[1em]
  \expl={Definition of \succFst}\\
  \all{(\lambda x^{Nat\times Nat}.}\\
  &\m{49}{\mu x_1^{\sms{\cnat 3,\cnat 4,Nat}\Rightarrow\cnat 1}.}\\
  &&\m{48}{\lambda n^{\cnat 3}.\lambda m^{\cnat 4}.\lambda k^{Nat}.}\\
  &&&\m{47}{\ifZ~(n\estr 3)}\\
  &&&&\m{46}{\canon 01}\\
  &&&&\m{46}{({\ifEq}~(m\estr 4)~(\suc k)}\\
  &&&&&\m{45}{\underline{\pi_{\cnat 1}}}\\
  &&&&&&\m{44}{\underline{(\lambda x^{Nat\times Nat}. (\suc(\fst(x)), \snd(x)))}}\\
  &&&&&&&\m{43}{\underline{\lbrack x_1\canon{\pred{(n\estr 3)}}3m0,R_2\canon{\pred{(n\estr 3)}}3m0\rbrack}}\\
  &&&&&\m{45}{(x_1\canon{\pred{(n\estr 3)}}3m(\suc k))}\\
  &&&&\m{46}{)}\\
  &&\m{48}{)}\\
  &\m{49}{\canon{\fst(x)}3}\\
  &\m{49}{\canon{\snd(x)}4}\\
  &\m{49}{0}\\
  \all{)\estr 1}\\[1em]
  \expl\to{Rule comm$_{ee}$}\\
  \all{(\lambda x^{Nat\times Nat}.}\\
  &\m{49}{\mu x_1^{\sms{\cnat 3,\cnat 4,Nat}\Rightarrow\cnat 1}.}\\
  &&\m{48}{\lambda n^{\cnat 3}.\lambda m^{\cnat 4}.\lambda k^{Nat}.}\\
  &&&\m{47}{\ifZ~(n\estr 3)}\\
  &&&&\m{46}{\canon 01}\\
  &&&&\m{46}{({\ifEq}~(m\estr 4)~(\suc k)}\\
  &&&&&\m{45}{\underline{\pi_{Nat\times Nat\Rightarrow\cnat 1}}}\\
  &&&&&&\m{44}{(\underline{\lambda x^{Nat\times Nat}}. (\suc(\fst(x)), \snd(x)))}\\
  &&&&&\m{45}{\lbrack x_1\canon{\pred{(n\estr 3)}}3m0,R_2\canon{\pred{(n\estr 3)}}3m0\rbrack}\\
  &&&&&\m{45}{(x_1\canon{\pred{(n\estr 3)}}3m(\suc k))}\\
  &&&&\m{46}{)}\\
  &&\m{48}{)}\\
  &\m{49}{\canon{\fst(x)}3}\\
  &\m{49}{\canon{\snd(x)}4}\\
  &\m{49}{0}\\
  \all{)\estr 1}\\[1em]
  \expl\to{Rule comm$_{ei}$}\\
  \all{(\lambda x^{Nat\times Nat}.}\\
  &\m{49}{\mu x_1^{\sms{\cnat 3,\cnat 4,Nat}\Rightarrow\cnat 1}.}\\
  &&\m{48}{\lambda n^{\cnat 3}.\lambda m^{\cnat 4}.\lambda k^{Nat}.}\\
  &&&\m{47}{\ifZ~(n\estr 3)}\\
  &&&&\m{46}{\canon 01}\\
  &&&&\m{46}{({\ifEq}~(m\estr 4)~(\suc k)}\\
  &&&&&\m{45}{(\lambda x^{Nat\times Nat}.}\\ 
  &&&&&&\m{44}{\underline{\pi_{\cnat 1}(\suc(\fst(x)), \snd(x))})}\\
  &&&&&\m{45}{\lbrack x_1\canon{\pred{(n\estr 3)}}3m0,R_2\canon{\pred{(n\estr 3)}}3m0\rbrack}\\
  &&&&&\m{45}{(x_1\canon{\pred{(n\estr 3)}}3m(\suc k))}\\
  &&&&\m{46}{)}\\
  &&\m{48}{)}\\
  &\m{49}{\canon{\fst(x)}3}\\
  &\m{49}{\canon{\snd(x)}4}\\
  &\m{49}{0}\\
  \all{)\estr 1}\\[1em]
  \expl{\to^2}{Rules simpl and proj}\\
  \all{(\lambda x^{Nat\times Nat}.}\\
  &\m{49}{\mu x_1^{\sms{\cnat 3,\cnat 4,Nat}\Rightarrow\cnat 1}.}\\
  &&\m{48}{\lambda n^{\cnat 3}.\lambda m^{\cnat 4}.\lambda k^{Nat}.}\\
  &&&\m{47}{\ifZ~(n\estr 3)}\\
  &&&&\m{46}{\canon 01}\\
  &&&&\m{46}{({\ifEq}~(m\estr 4)~(\suc k)}\\
  &&&&&\m{45}{\underline{(\lambda x^{Nat\times Nat}.\canon{\suc(\fst(x))}1)}}\\
  &&&&&\m{45}{\underline{\sms{x_1\canon{\pred{(n\estr 3)}}3m0,R_2\canon{\pred{(n\estr 3)}}3m0}}}\\
  &&&&&\m{45}{(x_1\canon{\pred{(n\estr 3)}}3m(\suc k))}\\
  &&&&\m{46}{)}\\
  &&\m{48}{)}\\
  &\m{49}{\canon{\fst(x)}3}\\
  &\m{49}{\canon{\snd(x)}4}\\
  &\m{49}{0}\\
  \all{)\estr 1}\\[1em]
  \expl\to{Rule $\beta$}\\
  \all{(\lambda x^{Nat\times Nat}.}\\
  &\m{49}{(\mu x_1^{\sms{\cnat 3,\cnat 4,Nat}\Rightarrow\cnat 1}.}\\
  &&\m{48}{\lambda n^{\cnat 3}.\lambda m^{\cnat 4}.\lambda k^{Nat}.}\\
  &&&\m{47}{\ifZ~(n\estr 3)}\\
  &&&&\m{46}{\canon 01}\\
  &&&&\m{46}{({\ifEq}~(m\estr 4)~(\suc k)}\\
  &&&&&\m{45}{\lbrack\suc(\underline{\fst}}\\
  &&&&&&\m{44}{\underline{\sms{x_1\canon{\pred{(n\estr 3)}}3m0,R_2\canon{\pred{(n\estr 3)}}3m0}}}\\
  &&&&&\m{45}{)\rbrack^{\bnum 1}}\\
  &&&&&\m{45}{(x_1\canon{\pred{(n\estr 3)}}3m(\suc k))}\\
  &&&&\m{46}{)}\\
  &&\m{48}{)}\\
  &\m{49}{\canon{\fst(x)}3}\\
  &\m{49}{\canon{\snd(x)}4}\\
  &\m{49}{0}\\
  \all{)\estr 1}\\[1em]
  \expl{\to^2}{Rules simpl and proj}\\
  \all{(\lambda x^{Nat\times Nat}.}\\
  &\m{49}{(\mu x_1^{\sms{\cnat 3,\cnat 4,Nat}\Rightarrow\cnat 1}.}\\
  &&\m{48}{\lambda n^{\cnat 3}.\lambda m^{\cnat 4}.\lambda k^{Nat}.}\\
  &&&\m{47}{\ifZ~(n\estr 3)}\\
  &&&&\m{46}{\canon 01}\\
  &&&&\m{46}{({\ifEq}~(m\estr 4)~(\suc k)}\\
  &&&&&\m{45}{\canon{\suc((x_1\canon{\pred{(n\estr 3)}}3m0)\estr 1)}1}\\
  &&&&&\m{45}{(x_1\canon{\pred{(n\estr 3)}}3m(\suc k))}\\
  &&&&\m{46}{)}\\
  &&\m{48}{)}\\
  &\m{49}{\canon{\fst(x)}3}\\
  &\m{49}{\canon{\snd(x)}4}\\
  &\m{49}{0}\\
  \all{)\estr 1}\\
\end{supertabular}

\section{Trace of \s{even}}\label{ap:traceEven}
\begin{supertabular}{*{50}{p{2mm}@{}}}
  \all{\underline{\s{even}}}\\
  \expl={Definition of \s{even}}\\
  \all{\underline{\fstO{Nat\Rightarrow Nat}(\evenOdd)}}\\[1em]
  \expl={Definition of \fstO{}}\\
  \all{(\pi_{Nat\Rightarrow\cnat 1}(\underline{\evenOdd}))\estr 1}\\[1em]
  \expl={Definition of \evenOdd}\\
  \all{(\underline{\pi_{Nat\Rightarrow\cnat 1}}}\\
  &\m{49}{(\underline{\mu x^{Nat\Rightarrow Nat\times Nat}}.}\\
  &&\m{48}{\lambda n^{Nat}.\ifZ~n~(0,\suc 0)(\swap~(x~(\pred~n))))}\\
  \all{)\estr 1}\\[1em]
  \expl\to{Rule comm$_\mu$}\\
  \all{(\mu x_1^{Nat\Rightarrow\cnat 1}.\underline{\pi_{Nat\Rightarrow\cnat 1}}}\\
  &\m{49}{(\underline{\lambda n^{Nat}}.\ifZ~n~(0,\suc 0)}\\
  &&&\m{47}{(\swap}\\
  &&&&\m{46}{(\lbrack x_1,}\\
  &&&&&\m{45}{\mu x_2^{Nat\Rightarrow\cnat 2}.\underline{\pi_{Nat\Rightarrow\cnat 2}}}\\
  &&&&&&\m{44}{(\underline{\lambda m^{Nat}}.}\\
  &&&&&&&\m{43}{\ifZ~m~(0,\suc 0)(\swap(\sms{x_1,x_2}(\pred~m))))\rbrack}\\
  &&&&&\m{45}{(\pred~n)}\\
  &&&&\m{46}{)}\\
  &&&\m{47}{)}\\
  &\m{49}{)}\\
  \all{)\estr 1}\\[1em]
  \expl{\to^2}{Rule comm$_{ei}$ ($\times 2$)}\\
  \all{(\mu x_1^{Nat\Rightarrow\cnat 1}.\lambda n^{Nat}.\underline{\pi_{\cnat 1}}}\\
  &\m{49}{(\underline{\ifZ}~n~(0,\suc 0)}\\
  &&&\m{47}{(\swap}\\
  &&&&\m{46}{(\lbrack x_1,}\\
  &&&&&\m{45}{\mu x_2^{Nat\Rightarrow\cnat 2}.\lambda m^{Nat}.\underline{\pi_{\cnat 2}}}\\
  &&&&&&\m{44}{(\underline{\ifZ}~m~(0,\suc 0)(\swap(\sms{x_1,x_2}(\pred~m))))\rbrack}\\
  &&&&&\m{45}{(\pred~n)}\\
  &&&&\m{46}{)}\\
  &&&\m{47}{)}\\
  &\m{49}{)}\\
  \all{)\estr 1}\\[1em]
  \expl{\to^2}{Rule comm$_{\textsl{ifZ}}$}\\
  \all{(\mu x_1^{Nat\Rightarrow\cnat 1}.\lambda n^{Nat}.}\\
  &\m{49}{(\ifZ~n~\underline{\pi_{\cnat 1}(0,\suc 0)}}\\
  &&\m{48}{\pi_{\cnat 1}}\\
  &&&\m{47}{(\swap}\\
  &&&&\m{46}{(\lbrack x_1,}\\
  &&&&&\m{45}{\mu x_2^{Nat\Rightarrow\cnat 2}.\lambda m^{Nat}.}\\
  &&&&&&\m{44}{(\underline{\ifZ}~m~\underline{\pi_{\cnat 2}(0,\suc 0)}}\\
  &&&&&&&\m{43}{\pi_{\cnat 2}(\swap(\sms{x_1,x_2}(\pred~m))))\rbrack}\\
  &&&&&\m{45}{(\pred~n)}\\
  &&&&\m{46}{)}\\
  &&&\m{47}{)}\\
  &\m{49}{)}\\
  \all{)\estr 1}\\[1em]
  \expl{\to^4}{Rules simp and proj}\\
  \all{(\mu x_1^{Nat\Rightarrow\cnat 1}.\lambda n^{Nat}.}\\
  &\m{49}{(\ifZ~n~\canon 01}\\
  &&\m{48}{\pi_{\cnat 1}}\\
  &&&\m{47}{(\swap}\\
  &&&&\m{46}{(\lbrack x_1,}\\
  &&&&&\m{45}{\mu x_2^{Nat\Rightarrow\cnat 2}.\lambda m^{Nat}.}\\
  &&&&&&\m{44}{(\ifZ~m~\canon{\suc 0}2}\\
  &&&&&&&\m{43}{\pi_{\cnat 2}(\underline{\swap}(\sms{x_1,x_2}(\pred~m))))\rbrack}\\
  &&&&&\m{45}{(\pred~n)}\\
  &&&&\m{46}{)}\\
  &&&\m{47}{)}\\
  &\m{49}{)}\\
  \all{)\estr 1}\\[1em]
  \expl={Definition of \swap}\\
  \all{(\mu x_1^{Nat\Rightarrow\cnat 1}.\lambda n^{Nat}.}\\
  &\m{49}{(\ifZ~n~\canon 01}\\
  &&\m{48}{\pi_{\cnat 1}}\\
  &&&\m{47}{(\swap}\\
  &&&&\m{46}{(\lbrack x_1,}\\
  &&&&&\m{45}{\mu x_2^{Nat\Rightarrow\cnat 2}.\lambda m^{Nat}.}\\
  &&&&&&\m{44}{(\ifZ~m~\canon{\suc 0}2}\\
  &&&&&&&\m{43}{\pi_{\cnat 2}}\\
  &&&&&&&&\m{42}{(\lambda x^{Nat\times Nat}.(\snd x,\fst x)}\\
  &&&&&&&&\m{42}{(\underline{\sms{x_1,x_2}(\pred~m)})}\\
  &&&&&&&&\m{42}{)}\\
  &&&&&&\m{44}{)}\\
  &&&&\m{46}{\rbrack}\\
  &&&&\m{46}{(\pred~n)}\\
  &&&&\m{46}{)}\\
  &&&\m{47}{)}\\
  &\m{49}{)}\\
  \all{)\estr 1}\\[1em]
  \expl\to{Rule dist$_i$}\\
  \all{(\mu x_1^{Nat\Rightarrow\cnat 1}.\lambda n^{Nat}.}\\
  &\m{49}{(\ifZ~n~\canon 01}\\
  &&\m{48}{\pi_{\cnat 1}}\\
  &&&\m{47}{(\swap}\\
  &&&&\m{46}{(\lbrack x_1,}\\
  &&&&&\m{45}{\mu x_2^{Nat\Rightarrow\cnat 2}.\lambda m^{Nat}.}\\
  &&&&&&\m{44}{(\ifZ~m~\canon{\suc 0}2}\\
  &&&&&&&\m{43}{\pi_{\cnat 2}}\\
  &&&&&&&&\m{42}{\underline{(\lambda x^{Nat\times Nat}.(\snd x,\fst x)}}\\
  &&&&&&&&\m{42}{\underline{(\sms{x_1(\pred~m),x_2(\pred~m)})}}\\
  &&&&&&&&\m{42}{)}\\
  &&&&&&\m{44}{)}\\
  &&&&\m{46}{\rbrack}\\
  &&&&\m{46}{(\pred~n)}\\
  &&&&\m{46}{)}\\
  &&&\m{47}{)}\\
  &\m{49}{)}\\
  \all{)\estr 1}\\[1em]
  \expl\to{Rule $\beta$}\\
  \all{(\mu x_1^{Nat\Rightarrow\cnat 1}.\lambda n^{Nat}.}\\
  &\m{49}{(\ifZ~n~\canon 01}\\
  &&\m{48}{\pi_{\cnat 1}}\\
  &&&\m{47}{(\swap}\\
  &&&&\m{46}{(\lbrack x_1,}\\
  &&&&&\m{45}{\mu x_2^{Nat\Rightarrow\cnat 2}.\lambda m^{Nat}.}\\
  &&&&&&\m{44}{(\ifZ~m~\canon{\suc 0}2}\\
  &&&&&&&\m{43}{\underline{\pi_{\cnat 2}}}\\
  &&&&&&&&\m{42}{\underline{(\snd (\sms{x_1(\pred~m),x_2(\pred~m)}),}}\\
  &&&&&&&&\m{42}{\underline{\fst (\sms{x_1(\pred~m),x_2(\pred~m)}))}}\\
  &&&&&&\m{44}{)}\\
  &&&&\m{46}{\rbrack}\\
  &&&&\m{46}{(\pred~n)}\\
  &&&&\m{46}{)}\\
  &&&\m{47}{)}\\
  &\m{49}{)}\\
  \all{)\estr 1}\\[1em]
  \expl{\to^2}{Rules simp and proj}\\
  \all{(\mu x_1^{Nat\Rightarrow\cnat 1}.\lambda n^{Nat}.}\\
  &\m{49}{(\ifZ~n~\canon 01}\\
  &&\m{48}{\pi_{\cnat 1}}\\
  &&&\m{47}{(\swap}\\
  &&&&\m{46}{(\lbrack x_1,}\\
  &&&&&\m{45}{\mu x_2^{Nat\Rightarrow\cnat 2}.\lambda m^{Nat}.}\\
  &&&&&&\m{44}{(\ifZ~m~\canon{\suc 0}2}\\
  &&&&&&&\m{43}{\canon{\underline{\fst (\sms{x_1(\pred~m),x_2(\pred~m)})}}2}\\
  &&&&&&\m{44}{)}\\
  &&&&\m{46}{\rbrack}\\
  &&&&\m{46}{(\pred~n)}\\
  &&&&\m{46}{)}\\
  &&&\m{47}{)}\\
  &\m{49}{)}\\
  \all{)\estr 1}\\[1em]
  \expl{\to^2}{Rules simp and proj}\\
  \all{(\mu x_1^{Nat\Rightarrow\cnat 1}.\lambda n^{Nat}.}\\
  &\m{49}{(\ifZ~n~\canon 01}\\
  &&\m{48}{\pi_{\cnat 1}}\\
  &&&\m{47}{(\underline{\swap}}\\
  &&&&\m{46}{(\lbrack x_1,}\\
  &&&&&\m{45}{\mu x_2^{Nat\Rightarrow\cnat 2}.\lambda m^{Nat}.}\\
  &&&&&&\m{44}{(\ifZ~m~\canon{\suc 0}2~\canon{(x_1(\pred~m)\estr 1)}2)}\\
  &&&&\m{46}{\rbrack}\\
  &&&&\m{46}{(\pred~n)}\\
  &&&&\m{46}{)}\\
  &&&\m{47}{)}\\
  &\m{49}{)}\\
  \all{)\estr 1}\\[1em]
  \expl={Definition of \swap}\\
  \all{(\mu x_1^{Nat\Rightarrow\cnat 1}.\lambda n^{Nat}.}\\
  &\m{49}{(\ifZ~n~\canon 01}\\
  &&\m{48}{\underline{\pi_{\cnat 1}}}\\
  &&&\m{47}{(\underline{\lambda x^{Nat\times Nat}.(\snd x,\fst x)}}\\
  &&&&\m{46}{(\underline{\lbrack x_1,}}\\
  &&&&&\m{45}{\underline{\mu x_2^{Nat\Rightarrow\cnat 2}.\lambda m^{Nat}.}}\\
  &&&&&&\m{44}{\underline{(\ifZ~m~\canon{\suc 0}2~\canon{(x_1(\pred~m)\estr 1)}2)}}\\
  &&&&\m{46}{\underline{\rbrack}}\\
  &&&&\m{46}{\underline{(\pred~n)}}\\
  &&&&\m{46}{)}\\
  &&&\m{47}{)}\\
  &\m{49}{)}\\
  \all{)\estr 1}\\[1em]
  \expl\to{Rule comm$_{ee}$}\\
  \all{(\mu x_1^{Nat\Rightarrow\cnat 1}.\lambda n^{Nat}.}\\
  &\m{49}{(\ifZ~n~\canon 01}\\
  &&\m{48}{\underline{\pi_{Nat\times Nat\Rightarrow\cnat 1}}(\underline{\lambda x^{Nat\times Nat}}.(\snd x,\fst x))}\\
  &&\m{48}{(\lbrack x_1,}\\
  &&&\m{47}{\mu x_2^{Nat\Rightarrow\cnat 2}.\lambda m^{Nat}.}\\
  &&&&\m{46}{(\ifZ~m~\canon{\suc 0}2~\canon{(x_1(\pred~m)\estr 1)}2)}\\
  &&\m{48}{\rbrack}\\
  &&\m{48}{(\pred~n)}\\
  &&\m{48}{)}\\
  &\m{49}{)}\\
  \all{)\estr 1}\\[1em]
  \expl\to{Rule comm$_{ei}$}\\
  \all{(\mu x_1^{Nat\Rightarrow\cnat 1}.\lambda n^{Nat}.}\\
  &\m{49}{(\ifZ~n~\canon 01}\\
  &&\m{48}{(\lambda x^{Nat\times Nat}.\underline{\pi_{\cnat 1}(\snd x,\fst x)})}\\
  &&\m{48}{(\lbrack x_1,}\\
  &&&\m{47}{\mu x_2^{Nat\Rightarrow\cnat 2}.\lambda m^{Nat}.}\\
  &&&&\m{46}{(\ifZ~m~\canon{\suc 0}2~\canon{(x_1(\pred~m)\estr 1)}2)}\\
  &&\m{48}{\rbrack}\\
  &&\m{48}{(\pred~n)}\\
  &&\m{48}{)}\\
  &\m{49}{)}\\
  \all{)\estr 1}\\[1em]
  \expl{\to^2}{Rules simp and proj}\\
  \all{(\mu x_1^{Nat\Rightarrow\cnat 1}.\lambda n^{Nat}.}\\
  &\m{49}{(\ifZ~n~\canon 01}\\
  &&\m{48}{(\lambda x^{Nat\times Nat}.\canon{\snd x}1)}\\
  &&\m{48}{(\underline{\lbrack x_1,}}\\
  &&&\m{47}{\underline{\mu x_2^{Nat\Rightarrow\cnat 2}.\lambda m^{Nat}.}}\\
  &&&&\m{46}{\underline{(\ifZ~m~\canon{\suc 0}2~\canon{(x_1(\pred~m)\estr 1)}2)}}\\
  &&\m{48}{\underline{\rbrack}}\\
  &&\m{48}{\underline{(\pred~n)}}\\
  &&\m{48}{)}\\
  &\m{49}{)}\\
  \all{)\estr 1}\\[1em]
  \expl\to{Rule dist$_i$}\\
  \all{(\mu x_1^{Nat\Rightarrow\cnat 1}.\lambda n^{Nat}.}\\
  &\m{49}{(\ifZ~n~\canon 01}\\
  &&\m{48}{\underline{(\lambda x^{Nat\times Nat}.\canon{\snd x}1)}}\\
  &&\m{48}{\underline{\lbrack x_1(\pred~n),}}\\
  &&&\m{47}{\underline{\mu x_2^{Nat\Rightarrow\cnat 2}.(\lambda m^{Nat}.}}\\
  &&&&\m{46}{\underline{(\ifZ~m~\canon{\suc 0}2~\canon{(x_1(\pred~m)\estr 1)}2))}}\\
  &&&\m{47}{\underline{(\pred~n)}}\\
  &&\m{48}{\underline{\rbrack}}\\
  &\m{49}{)}\\
  \all{)\estr 1}\\[1em]
  \expl\to{Rule $\beta$}\\
  \all{(\mu x_1^{Nat\Rightarrow\cnat 1}.\lambda n^{Nat}.}\\
  &\m{49}{(\ifZ~n~\canon 01}\\
  &&\m{48}{\underline{\lbrack \snd}}\\
  &&&\m{47}{\underline{\lbrack x_1(\pred~n),}}\\
  &&&\m{47}{\underline{\mu x_2^{Nat\Rightarrow\cnat 2}.(\lambda m^{Nat}.}}\\
  &&&&\m{46}{\underline{(\ifZ~m~\canon{\suc 0}2~\canon{(x_1(\pred~m)\estr 1)}2))
  (\pred~n)
  \rbrack
  \rbrack^{\bnum 1}}}\\
  &\m{49}{)}\\
  \all{)\estr 1}\\[1em]
  \expl{\to^2}{Rules simpl and proj}\\
  \all{(\mu x_1^{Nat\Rightarrow\cnat 1}.\lambda n^{Nat}.}\\
  &\m{49}{(\ifZ~n~\canon 01}\\
  &&\m{48}{\lbrack(\mu x_2^{Nat\Rightarrow\cnat 2}.\underline{(\lambda m^{Nat}.}}\\
  &&&\m{47}{\underline{(\ifZ~m~\canon{\suc 0}2~\canon{(x_1(\pred~m)\estr 1)}2))}}\\
  &&\m{48}{\underline{(\pred~n))}\estr 2\rbrack^{\bnum 1}}\\
  &\m{49}{)}\\
  \all{)\estr 1}\\[1em]
  \expl\to{Rule $\beta$}\\
  \all{(\mu x_1^{Nat\Rightarrow\cnat 1}.\lambda n^{Nat}.}\\
  &\m{49}{(\ifZ~n~\canon 01}\\
  &&\m{48}{{\lbrack}({\mu x_2^{Nat\Rightarrow\cnat 2}.}}\\
  &&&\m{47}{{(\ifZ~(\pred~n)}}\\
  &&&&\m{46}{{\canon{\suc 0}2}}\\
  &&&&\m{46}{{\canon{(x_1(\pred~(\pred~n))\estr 1)}2)}}\\
  &&\m{48}{)\estr 2\rbrack^{\bnum 1}}\\
  &\m{49}{)}\\
  \all{)\estr 1}\\
\end{supertabular}

\end{document}